\documentstyle[12pt,epsf]{article}

\newcommand{\PT}{{\mbox{$\not\hspace{-.55ex}{P}_{\rm t}$}}}

\def\3{\ss}                                                                                        
\begin{document}                                                                                   
                    
\title{{\bf
\vspace*{1cm}
A Search for Excited Fermions in $e^+p$ Collisions at HERA
\vspace*{2cm}
}}
\author{ZEUS Collaboration}
\date{ }
\maketitle
\vspace{3cm}

\begin{abstract}
\noindent
Using the ZEUS detector at HERA,
we have searched for heavy excited states of electrons, neutrinos, and
quarks in $e^+p$ collisions at a center-of-mass energy of
300~GeV. With an integrated luminosity of 9.4~pb$^{-1}$, no evidence
was found for 
electroweak production and decay of such states. Limits on the
production cross section times branching ratio and on the characteristic 
couplings, $f/\Lambda$, are derived for masses up to 250~GeV. For the
particular choice $f/\Lambda = 1/M_{f^*}$, we exclude at the 95\% confidence
level excited electrons with mass between 30 and 200~GeV,
excited electron neutrinos with mass between  40 and 96~GeV, 
and quarks excited electroweakly with mass between 40 and 169~GeV.
\end{abstract}

\vskip -18.6cm
\centerline{{\tt DESY 97-112}\hfill{\tt ISSN 0418-9833}}
\centerline{{\tt June 1997}\hfill}

\pagestyle{plain}
                    
\thispagestyle{empty}

\newpage

\pagenumbering{Roman}                                                                              
                                                   %
\begin{center}                                                                                     
{                      \Large  The ZEUS Collaboration              }                               
\end{center}                                                                                       
{\small                                                   %
  J.~Breitweg,                                                                                     
  M.~Derrick,                                                                                      
  D.~Krakauer,                                                                                     
  S.~Magill,                                                                                       
  D.~Mikunas,                                                                                      
  B.~Musgrave,                                                                                     
  J.~Repond,                                                                                       
  R.~Stanek,                                                                                       
  R.L.~Talaga,                                                                                     
  R.~Yoshida,                                                                                      
  H.~Zhang  \\                                                                                     
 {\it Argonne National Laboratory, Argonne, IL, USA}~$^{p}$
% ...
\newpage                                                                                           
$^{\    1}$ also at IROE Florence, Italy \\                                                        
% ...
$^{  23}$ now at ATLAS Collaboration, Univ. of Munich \\                                           
% ...
                                                           %
                                                           %
\newpage   
                                                           %
                                                           %
\begin{tabular}[h]{rp{14cm}}                                                                       
$^{a}$ &  supported by the Natural Sciences and Engineering Research                               
          Council of Canada (NSERC)  \\                                                            
$^{b}$ &  supported by the FCAR of Qu\'ebec, Canada  \\                                            
$^{c}$ &  supported by the German Federal Ministry for Education and                               
          Science, Research and Technology (BMBF), under contract                                  
          numbers 057BN19P, 057FR19P, 057HH19P, 057HH29P, 057SI75I \\                              
$^{d}$ &  supported by the MINERVA Gesellschaft f\"ur Forschung GmbH,                              
          the German Israeli Foundation, and the U.S.-Israel Binational                            
          Science Foundation \\                                                                    
$^{e}$ &  supported by the German Israeli Foundation, and                                          
          by the Israel Science Foundation                                                         
  \\                                                                                               
$^{f}$ &  supported by the Italian National Institute for Nuclear Physics                          
          (INFN) \\                                                                                
$^{g}$ &  supported by the Japanese Ministry of Education, Science and                             
          Culture (the Monbusho) and its grants for Scientific Research \\                         
$^{h}$ &  supported by the Korean Ministry of Education and Korea Science                          
          and Engineering Foundation  \\                                                           
$^{i}$ &  supported by the Netherlands Foundation for Research on                                  
          Matter (FOM) \\                                                                          
$^{j}$ &  supported by the Polish State Committee for Scientific                                   
          Research, grant No.~115/E-343/SPUB/P03/002/97, 2P03B10512,                               
          2P03B10612, 2P03B14212, 2P03B10412 \\                                                    
$^{k}$ &  supported by the Polish State Committee for Scientific                                   
          Research (grant No. 2P03B08308) and Foundation for                                       
          Polish-German Collaboration  \\                                                          
$^{l}$ &  partially supported by the German Federal Ministry for                                   
          Education and Science, Research and Technology (BMBF)  \\                                
$^{m}$ &  supported by the German Federal Ministry for Education and                               
          Science, Research and Technology (BMBF), and the Fund of                                 
          Fundamental Research of Russian Ministry of Science and                                  
          Education and by INTAS-Grant No. 93-63 \\                                                
$^{n}$ &  supported by the Spanish Ministry of Education                                           
          and Science through funds provided by CICYT \\                                           
$^{o}$ &  supported by the Particle Physics and                                                    
          Astronomy Research Council \\                                                            
$^{p}$ &  supported by the US Department of Energy \\                                              
$^{q}$ &  supported by the US National Science Foundation \\                                       
\end{tabular}                                                                                      
}                                                           %
                                                       %
\newpage \pagestyle{plain} \pagenumbering{arabic}

\section{Introduction}

\label{sec:intro}

The existence of excited leptons or quarks would provide clear
evidence for fermion substructure. Any theory of
compositeness, however, would require a low energy limit which respects the
symmetries of the Standard Model within the bounds provided by present
experimental data. At the HERA electron-proton collider, single excited
electrons ($e^{*}$) and quarks ($q^{*}$) could be produced by $t$-channel $
\gamma /Z^0$ boson exchange, and excited neutrinos ($\nu^{*}$) could
be produced by $t$-channel $W$ boson exchange. These mechanisms are depicted
in Fig.~\ref{fig:feyn}.

We report on a search for resonances with masses above 30~GeV, produced in 
$e^{+}p$ collisions at $\sqrt{s}=300$~GeV, decaying to a light fermion
through electroweak mechanisms. We interpret the results in the
context of excited fermion ($f^*$) decays.
The initial state positron implies that HERA would produce excited
anti-leptons. To be succinct, the distinction between particles and
anti-particles will be dropped in this paper. For example, ``electron'' will
be used generically to refer to both $e^{-}$ and $e^{+}$.
The data sample, collected by ZEUS during
the years 1994--1995, corresponds to an integrated luminosity of 9.4~pb$^{-1}$.
This represents a 17-fold statistical increase over our previously
published \cite{ZEUS} limits from $e^{-}p$ collisions. A search based on
2.75~pb$^{-1}$ of $e^{+}p$ data has been reported recently by the H1
Collaboration \cite{H1}.

This paper is organized as follows. In the next section, we review the
phenomenology of excited fermion production; in Section~\ref{sec:limits} we
review the experimental constraints from direct searches at LEP and the
Tevatron; and in Section~\ref{sec:setup} we describe the ZEUS detector, trigger
configuration, and Monte Carlo simulations used to calculate acceptances and
backgrounds. Section~\ref{sec:kinematics} describes the electron and photon
identification criteria and the event kinematics determined from the
calorimeter. Section~\ref{sec:analysis} describes the selection criteria for
the several $f^{*}$ decay modes, followed in Section \ref{sec:systematics}
by an estimate of the systematic uncertainties associated with these
selections. In Section~\ref{sec:results}, we present the results on $f^{*}$
production cross sections and couplings, and compare them with prior
limits (an Appendix is included describing the details of our upper limit
derivation). Finally, Section~\ref{sec:conclusion} summarizes the
conclusions.

\section{Excited Fermion Production Models}

\label{sec:theory}

To maintain generality in describing $f^{*}$ production and decay,
without specifying the dynamics of the compositeness, it has been
conventional to use the phenomenological Lagrangian of Hagiwara, Komamiya,
and Zeppenfeld \cite{Hagiwara} describing the magnetic transitions between
ordinary fermions ($f$) and spin~$\frac 12$ excited states: 
\begin{equation}
{\cal L}_{f f^{*}}\;=\;{\frac e\Lambda }\,\sum_{V=\gamma ,Z,W}\,
\overline{f}^{*}\,\sigma ^{\mu \nu }\,(c_{Vf^{*}f}-d_{Vf^{*}f}\,\gamma
^5\,)\,f\,\partial _\mu V_\nu \;+\;h.c.  \label{eq:Hagiwara}
\end{equation}
Here $\Lambda $ is the compositeness scale, and $c_{Vf^{*}f}$ and 
$d_{Vf^{*}f}$ are coupling constants at the 
$f\leftrightarrow f^{*}$ transition vertex, 
labelled for each vector boson, $V$. 
Excitations of spin~$\frac 32$ excited states, which involve a
larger number of arbitrary parameters, are discussed in Ref.~\cite{Kuhn} but
will not be considered here.

The agreement between the precise measurements of electron/muon $g-2$ and
theoretical predictions implies that $|c_{\gamma f^{*}f}|=|d_{\gamma
f^{*}f}| $ for compositeness scales less than 10--100~TeV~\cite{Renard}. The
absence of electron and muon electric dipole moments forces $c_{\gamma
f^{*}f}$ and $d_{\gamma f^{*}f}$ to have the same phase. It is
customary to choose a model \cite{Hagiwara} which
couples left-handed fermions to right-handed excited
states, and in which the excited fermions form both left-
and right-handed weak isodoublets. 
The interaction Lagrangian, including excited leptons and
quarks \cite{Baur,Boudjema}, can be written: 
\begin{equation}
{\cal L}_{f f^{*}}\;=\;{\frac 1\Lambda }\;\overline{f}_{{\rm R}
}^{*}\;\sigma ^{\mu \nu }\;\Big[ gf{\frac{\tau ^{{\rm a}}}2}{\bf W}_{\mu \nu
}^{{\rm a}}+g^{\prime }f^{\prime }{\frac Y2}B_{\mu \nu }+g_{{\rm s}}f_{{\rm s
}}{\frac{\lambda ^{{\rm a}}}2}{\bf G}_{\mu \nu }^{{\rm a}}\Big]\;f_{{\rm L}
}\;+h.c.  \label{eq:Baur}
\end{equation}
where ${\bf W}_{\mu \nu }^{{\rm a}}$, $B_{\mu \nu }$, and ${\bf G}_{\mu \nu
}^{{\rm a}}$ are the field-strength tensors of the $SU(2)_L$, $U(1)_Y$, and $
SU(3)_C$ gauge fields respectively; $\tau ^{{\rm a}}$, $Y$, and $\lambda ^{
{\rm a}}$ are the corresponding gauge group generators; and $g$, $
g^{\prime }$, and $g_{{\rm s}}$ are the corresponding gauge coupling
constants. The unknown parameters $f$, $f^{\prime }$, and $f_{{\rm s}}$
depend on the specific dynamics describing the compositeness. 
The electroweak coupling constants of Eq.~(\ref{eq:Hagiwara}), with
$c_{Vf^{*}f}=d_{Vf^{*}f}$, may be expressed\footnote{
These definitions for $f$, $f^{\prime }$, and $f_{{\rm s}}$ differ by a
factor of 2 from the convention used in Refs. \cite{Baur} and \cite{Boudjema}
.}: 
\begin{eqnarray}
c_{\gamma f^{*}f} &=&\frac 12(f\,T_3+f^{\prime }\,{\frac Y2})  \nonumber \\
c_{Zf^{*}f} &=&\frac 12(f\,T_3\cot \theta _{{\rm W}}-f^{\prime }\,{\frac Y2}
\tan \theta _{{\rm W}})  \label{eq:cgen} \\
c_{Wf^{*}f} &=&{\frac f{2\sqrt{2}\sin \theta _{{\rm W}}}} 
\nonumber 
\end{eqnarray}
where $T_3$ is the third component of the weak isospin, $Y$ is the weak
hypercharge ($-1$ for leptons and $\frac 13$ for quarks), and $\theta _{{\rm
W}}$ is the weak mixing angle. For specific assumptions relating $f$, 
$f^{\prime }$, and $f_{{\rm s}}$, the branching fractions are
known. Additionally, the cross sections
are described by a single parameter (e.g., $
f/\Lambda \,$) with dimension, GeV$^{-1}$.

\begin{figure}[p]
\centerline{\epsfbox{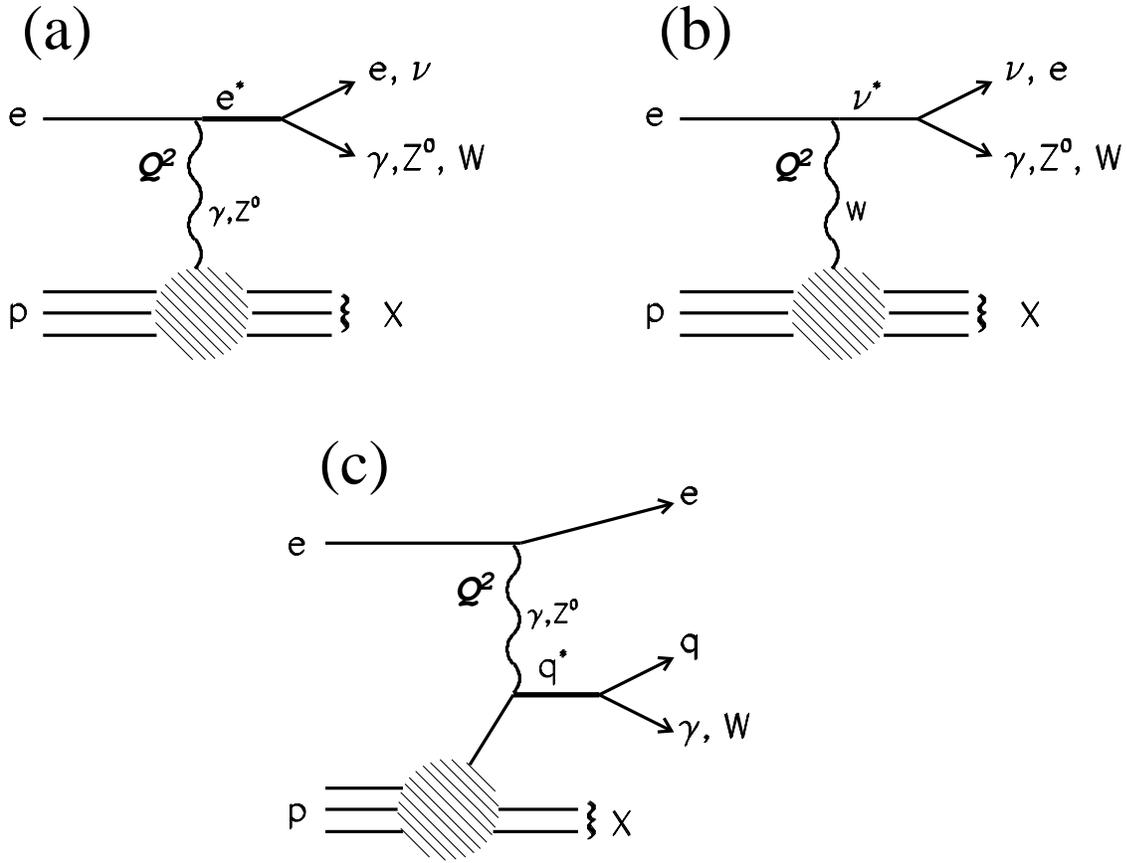}}
\caption{\label{fig:feyn} Diagrams for the production of
(a) excited electrons, (b) excited neutrinos, and (c) excited quarks
in $e\,p$ collisions.
Only those decay modes considered in this paper are shown.
}
\end{figure}

The cross section for $f^*$ production in $e\,p$ collisions has been
calculated \cite{Hagiwara,Boudjema}.
For excited electron ($f^{*}=e^{*} $) production, $t$-channel $\gamma$ 
exchange (Fig. \ref{fig:feyn}a)
is expected to dominate, with the elastic contribution ($X=p$) approximately
50\% of the cross section. Production of excited neutrinos 
($\nu_{\rm e}^{*}$) proceeds via $W$ exchange (Fig. \ref{fig:feyn}b),
for which the negative
square of the four-momentum transferred by the exchanged boson ($Q^2$) is
large, the cross section is small, and there is no elastic channel.
We also searched for excited quarks ($q^{*}$)
produced through electroweak couplings illustrated in Fig. \ref{fig:feyn}c.
This is complementary to $q^{*}$ searches at hadron colliders.

For the presumed large $f^{*}$ masses involved in all these searches,
a substantial 
fraction of the available center-of-mass energy is required to provide the
mass, $M_{f^{*}}$. HERA kinematics forces any $f^{*}$ with mass
larger than twice the electron beam energy to be boosted in the proton beam
direction with little transverse momentum.
The decay products of the $f^{*}$, on the
other hand, typically have large laboratory angles and large
transverse momenta. 

Produced $f^{*}$ decay to a light fermion and a gauge boson. The
partial electroweak decay widths for $f^{*}\rightarrow f\,V$ (with $V=\gamma 
$, $Z^0$, or $W$) are given by \cite{Boudjema} 
\begin{equation}
\Gamma (f^{*}\rightarrow fV)\;=\;\alpha \,{\frac{M_{f^{*}}^3}{\Lambda ^2}}
\,c_{Vf^{*}f}^2\,\Big(1-{\frac{M_V^2}{M_{f^{*}}^2}}\Big)^2\,\Big(1+{\frac{
M_V^2}{2M_{f^{*}}^2}}\Big)  \label{width}
\end{equation}
for the vector boson mass $M_V<M_{f^{*}}$.
As is conventional,
we choose $f=f^{\prime }$ in interpreting 
our search for $e^{*}$ decay  (for $M_{f^{*}}\gg M_V$, the $W$ decay mode for
excited electrons dominates with this choice).  The decay 
$\nu^*_{\rm e}\rightarrow \nu_{\rm e}\,\gamma$
is forbidden unless $f\neq f^{\prime }$. We choose  
$f=-f^{\prime}$ in the interpretation of our search for $\nu_{{\rm e}}^*$
decays. The partial width for $q^{*}$
decaying with gluon emission is obtained from (\ref{width}) by replacing the
electroweak coupling, $\alpha $, with $\frac 43\alpha _{{\rm s}}$,
where $\alpha _{{\rm s}}$ is the quark-gluon coupling. For the
assumption that $f_{{\rm s}}$ 
is comparable in magnitude to $f$ and $f^{\prime }$, the decay of excited
quarks into gluons dominates all other decays since $\alpha _{{\rm s}}$
is so much larger than $\alpha $. As discussed in Section~\ref{sec:limits},
Tevatron experiments limit $f_{{\rm s}}$, whereas for small $f_{{\rm s}}$,
the high mass electroweak limits provided here on 
$f$ and $f^{\prime }$ are unique.
For values of $M_{f^*}$ considered in this paper, the
total electroweak decay width is approximately 1~GeV or less.

\section{Direct Limits from $e^+\,e^-$ and $\overline{p}\,p$ Colliders}

\label{sec:limits}

Excited leptons and quarks also might be directly produced in $e^{+}e^{-}$
and $\overline{p}p$ collisions. LEP searches at $\sqrt{s}=130$--$140$~GeV 
\cite{LEPonefive} and at $\sqrt{s}=161$~GeV \cite{LEPtwo} rule out at the
95\% confidence level any excited lepton ($e^{*},\mu ^{*},\tau ^{*},\nu ^{*}$
) with mass below approximately 80~GeV. Since such excited fermions would
be pair produced, these limits 
are independent of the coupling parameters $f$
and $f^{\prime }$. Single excited leptons also can be produced at LEP
in a manner analogous to that at HERA: each of the four experiments set
upper limits at the 95\% confidence level on $f/\Lambda $ (assuming $
f=f^{\prime }$) at the level of (0.3--1.0)$\times 10^{-3}$~GeV$^{-1}$ for
excited electrons and ${\cal O}(10^{-2})$~GeV$^{-1}$ for all other lepton
species with masses up to nearly the LEP center-of-mass energy. The larger
center-of-mass energy at HERA allows for sensitivity
at higher masses.

The CDF experiment has searched for excited quarks produced in $\overline{p}
\,p$ collisions at $\sqrt{s}=1800$~GeV.  Initially, only decay modes where 
the excited quark decays
to a quark and photon or to a quark and $W$ boson decaying
leptonically \cite{CDFold} were considered.
Limits were set
at the 95\% confidence level on excited quarks in the mass
range 80--540~GeV under the assumptions $f=f^{\prime }=f_{{\rm s}}=\frac 12$
(their convention differs by a factor two from ours) and $\Lambda =M_{q^{*}}$.
More recently, the same group has looked for evidence in high transverse
energy jet samples of a resonance characteristic of excited
quarks decaying through gluon couplings \cite{CDFnew}, so that visible
signatures require only $f_s\neq 0$. No evidence of a resonance was seen,
permitting extension of the upper limit under the above assumptions 
to 760 GeV, except for a window between 570 and 580 GeV.
These results translate to an upper limit on $
f_{{\rm s}}/\Lambda $ of (0.5--1.0)$\times 10^{-3}$~GeV$^{-1}$
for the given mass range under the stated assumptions. 

It should be noted
that searches for excited quarks at HERA\ are complementary to those
at hadron colliders. All Tevatron limits assume coupling
through the gluon ($ f_s\neq 0$) whereas HERA limits on electroweak
couplings $f$ and $f^{\prime}\,$ are strongest
for $f_s=0$.

\section{Experimental Setup and Event Simulation}

\label{sec:setup}

The data used in this analysis were collected with the ZEUS detector.
The HERA beam energies were 27.5~GeV for positrons and
820~GeV for protons. 

\subsection{The ZEUS Detector}

\label{sec:zeus}

ZEUS is a nearly hermetic, multipurpose, magnetic detector
described in detail elsewhere \cite{DET,Detector}. The primary components
used in this analysis are the high resolution depleted-uranium scintillator
calorimeter, the central tracking detectors, and the luminosity detector.

The ZEUS coordinate system is right-handed with the $Z$ axis pointing in the
proton beam direction, referred to as the ``forward'' direction. The $X$
axis points horizontally toward the center of HERA, and the origin is at the
nominal interaction point. The polar angle $\theta $ is measured with
respect to the $Z$ direction, and the corresponding pseudorapidity is 
$\eta =-\ln \tan (\theta /2)$.

The compensating calorimeter \cite{CAL} consists of three parts: the forward
calorimeter (FCAL) covering $2.6^{\circ }<\theta <36.7^{\circ }$, the barrel
calorimeter (BCAL) covering $36.7^{\circ }<\theta <129.1^{\circ }$, and the
rear calorimeter (RCAL) covering $129.1^{\circ }<\theta <176.2^{\circ }$.
Each part of the calorimeter is subdivided longitudinally into one
electromagnetic section (EMC), and one or two hadronic sections (HAC) for
the RCAL or FCAL/BCAL, respectively. Each section is further subdivided
transversely into cells of $5\times 20$~cm$^2$ ($10\times 20$~cm$
^2 $ in RCAL) for the EMC sections and $20\times 20$~cm$^2$ for the HAC
sections. The calorimeter has an energy resolution of $\sigma _E/E=0.18/%
\sqrt{E({\rm GeV})}$ for electrons and $\sigma _E/E=0.35/\sqrt{E({\rm GeV})}$
for hadrons as measured under test beam conditions. The cell-to-cell
variations in the energy calibration are approximately $2\%$ for the EMC
cells and $3\%$ for HAC cells. The FCAL and BCAL energy scale calibrations
are presently known to $3\%$. The calorimeter provides time measurements
with resolution better than 1~ns for energy deposits above 4.5~GeV.

The tracking components used in this analysis are the vertex detector \cite
{VXD} and the central tracking detector \cite{CTD}, operating in a 1.43~T
solenoidal magnetic field. The central tracking detector is capable of
reconstructing the tracks over the region $15^{\circ }<\theta<165^{\circ }$
and also supplies a vertex measurement for each event. The transverse momentum
resolution for tracks intersecting all tracking 
layers is $\sigma (p_{{\rm t}})/p_{{\rm t}}=[0.005\,p_{{\rm t}}({\rm GeV}
)]\oplus 0.016$.

The luminosity was measured to a precision of about 1.5\% from the rate of
energetic bremsstrahlung photons produced in the process $e\,p\rightarrow
e\,p\,\gamma$. The photons are detected in a lead-scintillator calorimeter
\cite{lumi} placed at $Z=-107$~m.

\subsection{Trigger Configuration}

\label{sec:trigger}

Events were filtered online by a three-level trigger system~\cite{Detector}.
The trigger criteria used in this analysis relied primarily on the energies
measured in the calorimeter. The trigger decision was based on
electromagnetic energy, total transverse energy ($E_{{\rm t}}$), missing
transverse momentum ({\mbox{$\not\hspace{-.55ex}{P}_{\rm t}$}}),
$E-P_{\rm Z}$, and timing.
Thresholds on kinematic parameters, in general, 
were significantly below the corresponding offline cuts
described in Section~\ref{sec:analysis}.

\subsection{Monte Carlo Simulation}

\label{sec:mc}

The acceptance due to our selection criteria is calculated with
a Monte Carlo simulation of excited fermion production and decay.
This
analysis uses the code {\sc hexf} \cite{HEXF}, based on the models discussed
earlier \cite{Hagiwara,Baur,Boudjema}. Interference with
Standard Model processes 
is not considered. Initial state
radiation from the beam electron uses the
Weizs\"{a}cker-Williams approximation \cite{WW}. The hadronic final state is
simulated using the matrix element and parton shower approach of {\sc lepto}
\cite{LEPTO} for the hard QCD parton cascade, and {\sc jetset} \cite{JETSET}
for the soft hadronization. The MRSA\cite{MRSA} 
parton distribution set is used for inelastic scattering.

A particular choice of parameters must be made in the excited fermion
production model in order to determine the detector acceptance. We have
adopted the convention $f=f^{\prime }$ in the Lagrangian\footnote{
The value of $f_s$ is only relevant for the $q^{*}$ search, where we
describe the assumptions more fully.}~(\ref{eq:Baur}), except for the
decay $\nu^*_{\rm e}\rightarrow \nu_{\rm e}\,\gamma$
where we choose $f^{\prime }=-f$ to
achieve a non-zero coupling (note that the production cross section depends
only on $f$, however). The acceptance only weakly depends on these
assumptions.

Backgrounds from charged and neutral current Deep Inelastic Scattering (DIS)
are simulated using {\sc heracles} \cite{HERACLES}, including first-order
electroweak radiative corrections. The hadronic final states use 
{\sc lepto} and {\sc jetset} interfaced
to {\sc heracles} using {\sc django} \cite{DJANGO}.
Resolved and direct photoproduction backgrounds, including prompt photon
production, are simulated with the {\sc herwig} generator \cite{HERWIG}.
QED-Compton scattering events are generated with  
{\sc compton} \cite{COMPTON}. 
Finally, production of $W$ bosons, a potential background to rare
topologies, uses the {\sc epvec} generator \cite{EPVEC}.

All simulated events are passed through a detector simulation based on 
{\sc geant} \cite{GEANT}, 
incorporating our knowledge of the detector and trigger,
and subsequently processed by the same reconstruction 
and analysis programs used for data. 

\section{Electromagnetic Cluster Identification and Event Kinematics}

\label{sec:kinematics}

Excited fermions decaying to electrons or photons give rise to isolated
electromagnetic (EM) clusters in the calorimeter, identified by the
characteristically small longitudinal and lateral profiles. 
Pattern recognition combines a
neural network algorithm \cite{Sinistra} and a requirement limiting the shower
width along the direction of the 5~cm cell segmentation in the BCAL 
and FCAL (no cut is applied in the RCAL). The shower widths are 
required to be less than $4$~cm ($5$~cm) in the BCAL (FCAL).

EM clusters are said to be {\em isolated} if they 
satisfy a criterion based on the sum of the
transverse energy, $E_R$, within a cone of radius $R=\sqrt{(\Delta \eta
)^2+(\Delta \phi )^2}$ centered on the cluster, contributed by all
calorimeter cells {\em unassociated} with the EM cluster. Depending
on the desired degree of isolation, the choice for $
R$ is either 0.7 or 1.5, and the requirement is $E_R<2$~GeV.

The event selection for the excited fermion searches requires the following
global event quantities calculated from the event vertex and the 
calorimeter cell measurements: 
\begin{eqnarray}
E-P_{{\rm Z}} &=&{\ \sum_{\rm i}}\, \left( E_{{\rm i}}-P_{{\rm Z,i}}\right) 
\nonumber \\
E_{{\rm t}} &=&{\ \sum_{\rm i}}^{\prime}\, \sqrt{P_{{\rm X,i}}^2+P_{{\rm 
Y,i}}^2}
\label{eq:sumhad} \\
\not{\hspace{-.55ex}}{P}_{{\rm t}} &=&\sqrt{ \left( {\textstyle \sum_{\rm i}}\, P_{
{\rm X,i}}\right) ^2 + \left( {\textstyle \sum_{\rm i}}\, P_{{\rm Y,i}} \right) 
^2} \nonumber \\
M &=&\sqrt{\left( {\textstyle \sum_{\rm i}}^{\prime}\, E_{{\rm i}}\right) ^2 
 - \left( {\textstyle \sum_{\rm i}}^{\prime}\, P_{{\rm X,i}}\right) ^2 -\left( 
{\textstyle \sum_{\rm i}}^{\prime}\, P_{{\rm Y,i}}\right) ^2 -\left( 
{\textstyle \sum_{\rm i}}^{\prime}\, P_{{\rm Z,i}}\right) ^2}  \nonumber
\end{eqnarray}
where $E_i=\sqrt{P_{{\rm X,i}}^2 + P_{{\rm Y,i}}^2 + P_{{\rm Z,i}}^2}$ 
is the energy in the i$^{th}$ calorimeter cell.

$E_{{\rm t}}$\ is the total scalar sum of the transverse energy deposited in
the calorimeter, {\mbox{$\not\hspace{-.55ex}{P}_{\rm t}$}}\ is the missing
transverse momentum, and $M$ is the measured mass. 
Momentum conservation dictates that the longitudinal momentum variable, $
E-P_{{\rm Z}}$, equals twice the incident electron beam energy ($
2E_e=55$~GeV) for final states in which 
no energy escapes through the rear beam hole.

The unprimed sums run over all calorimeter cells with energy deposits above
a set threshold. The primed sums (in the calculations of $E_{{\rm t}}$ and $M$)
exclude the cells in the forward cone, $\theta <10^{\circ }$
(this reduces sensitivity to particles within the proton remnant).
Any of these quantities with the subscript ``had'' 
also exclude cells in the
calorimeter sum (double primed sum) 
which belong to clusters which are identified as candidate electrons
or photons from $f^*$ decay.
For  example, the azimuthal ($\phi _{{\rm 
had}}$) and polar ($\theta _{{\rm had}}$) angles respectively for the
hadronic system are defined:  
\begin{eqnarray}
\phi _{{\rm had}} &=&\arctan\,{\frac{{\sum_{\rm i}}^{\prime\prime }\,P_{{\rm Y,i}}}{{
\sum_{\rm i}}^{\prime\prime }\,P_{{\rm X,i}}}}  \label{eq-phihad} \\
\theta _{{\rm had}} &=&\arccos\,{\frac{{\sum_{\rm i}}^{\prime\prime }\,P_{{\rm Z,i}}}{
\sqrt{\left( {\textstyle \sum_{\rm i}}^{\prime\prime }\,P_{{\rm X,i}}\right) ^2+\left( {
\textstyle \sum_{\rm i}}^{\prime\prime }\,P_{{\rm Y,i}}\right) ^2+\left( {\textstyle 
\sum_{\rm i}}^{\prime\prime }\,P_{{\rm Z,i}}\right) ^2}}.}  \nonumber
\end{eqnarray}

\section{Event Selection}

\label{sec:analysis}

The processes of interest for production of the various $f^{*}$ are 
\begin{equation}
\begin{array}{l}
e+p\rightarrow e^{*}+X \\ 
e+p\rightarrow \nu ^{*}+X
\end{array}
\label{eq:enuprod}
\end{equation}
for excited leptons, and 
\begin{equation}
e+p\rightarrow e+q^{*}+X  \label{eq:qstprod}
\end{equation}
for excited quarks. Here $X$ represents the proton remnant (or proton
in the case of elastic $e^*$ production).
An important feature of these processes is
that reactions (\ref{eq:enuprod}) involve an excited lepton which
subsequently decays into at least one high transverse momentum lepton.
Reaction (\ref{eq:qstprod}), by contrast, involves a final state excited quark 
${\em and}$ a final state electron. This electron typically travels in the
direction of the electron beam and has low transverse momentum.

The searches are organized in this section by event topology.
We have sought the following
excited fermion decays:

\begin{enumerate}
\item  $e^{*}\rightarrow e\,\gamma $

\item  $e^{*}\rightarrow e\,Z^0\rightarrow e\,q\,\overline{q}$ ~~and~~ $\nu
_{{\rm e}}^{*}\rightarrow e\,W\rightarrow e\,q\,\overline{q}$

\item  $e^{*}\rightarrow \nu _{{\rm e}}\,W\rightarrow \nu _{{\rm e}}\,q\,
\overline{q}$ ~~and~~ $\nu _{{\rm e}}^{*}\rightarrow \nu _{{\rm e}
}\,Z^0\rightarrow \nu _{{\rm e}}\,q\,\overline{q}$

\item  $\nu _{{\rm e}}^{*}\rightarrow \nu _{{\rm e}}\,\gamma $

\item  $e^{*}\rightarrow e\,Z^0\rightarrow e\,\nu\,\overline{\nu }$ 
~~and~~ $e^{*}\rightarrow \nu _{{\rm e}}\,W\rightarrow e\,\nu _{
{\rm e}}\,\overline{\nu }_{{\rm e}}$

\item  $\nu _{{\rm e}}^{*}\rightarrow e\,W\rightarrow 
       e\,\overline{e}\,\nu _{{\rm e}}$

\item  $q^{*}\rightarrow q\,\gamma $

\item  $q^{*}\rightarrow q\,W\rightarrow q\,\overline{e}\,\nu _{{\rm e}}$
\end{enumerate}

The conclusions for each $f^{*}$ are discussed in Section~\ref{sec:results} and
summarized in Tab.~\ref{tab:effic}. Also shown in this table are the 
decay signature, typical
acceptance, number of observed events, and number of expected events from
background sources for each $f^{*}$ decay mode 
described above.

The following selection cuts, common to all channels, were applied to ensure
that the selected events are $e\,p$ interactions.

\begin{enumerate}
\item  Events are required to have a reconstructed vertex 
measured by the central tracking
detectors, and it must lie within 50~cm of the nominal collision point
along the $Z$ axis.\footnote{
The $Z$ vertex distribution 
is approximately Gaussian with an r.m.s.
spread of 12 cm, a consequence of the proton
bunch length.}

\item  The arrival times measured with the calorimeter 
are required to be
consistent with final state particles originating from the nominal
collision point. 

\item  Pattern recognition algorithms are applied to suppress 
non-$e\,p$ backgrounds  (cosmic rays, beam halo muons, $p$--gas
interactions, and photomultiplier sparks). In addition, each final
event sample is visually inspected, and a few remaining
events are removed which fall into one of these background categories. 

\end{enumerate}

In addition, since the decay products of heavy leptons would be
forward-going, 
the total calorimeter energy, $E_{{\rm rear}}$, deposited in
the cone $\theta >150^{\circ }$ is required to be less than 2~GeV.  
As discussed in Section~\ref{sec:qgam}, this cut is different for
excited quark searches as a scattered electron may be detected in the
rear direction.

\subsection{$e\,\gamma$ Resonance Search}

\label{sec:egam}

The signature for the decay 
$e^{*}\rightarrow e\,\gamma $ is the observation in the calorimeter 
of two isolated EM clusters.
QED-Compton scattering ($e\,p\rightarrow e\,\gamma \,X$)
has a similar topology, and thus forms a non-resonant background at 
predominantly low
electron-photon masses, $M_{e\gamma }$. The angular distribution is
different \cite{Hagiwara}, however, so the two may be statistically
separated. Additional backgrounds arise from neutral current DIS events 
with an isolated $\pi^0$.  Two-photon production of electron pairs
also produces two electromagnetic showers, albeit with tracks,
but is found to be negligible.

The following selections are required:

\begin{enumerate}
\item Each EM cluster is required to have a transverse energy
larger than 10~GeV, and each cluster must be isolated from hadronic activity
by an $\eta $--$\phi $ cone radius $R=0.7$ 
(see Section~\ref{sec:kinematics}).  We do not require a track to be
associated with either EM cluster.

\item Low $Q^2$ DIS events with a $\pi ^0$
in the forward region 
are suppressed by requiring that the
polar angle of each EM cluster be larger than $15^{\circ }$, or
that each EM cluster energy be larger than 32~GeV (the polar angle cut
alone would severely limit the acceptance for high mass excited electrons).

\item  The net $E-P_{{\rm Z}}$ 
of the two electromagnetic clusters 
must reflect longitudinal momentum conservation, so it is required to be
in the range
35 to 65~GeV. This limits any vertex mismeasurement as well as
initial state radiation.

\item  The transverse energy deposited in the calorimeter by the two
clusters must account for more than 70\% of the total transverse energy.
This reduces neutral current
DIS background.

\item  Events with more than three well-measured tracks in the central
tracking detector\footnote{
Such tracks must have transverse momentum $P_{{\rm t}}>0.2$~GeV and
$15^{\circ }<\theta <165^{\circ }$.} 
are rejected. This preferentially selects elastic $e^*$ production.

\end{enumerate}

The selection criteria accept $e^{*}$ produced with momentum transfers 
as large as $Q^2\approx 1000$~GeV$^2$, and so includes a large fraction of 
the expected production (with $f=f^{\prime}$).\footnote{
The model
of Ref.~\cite{Hagiwara} (with $f=f^{\prime }$) predicts that $e^{*}$ are
preferentially produced at low $Q^2$; approximately 60\% are produced
elastically or quasi-elastically.
Only for $f^{\prime }\approx -f$ in Eq.~(\ref{eq:cgen}) is the coupling to
photons suppressed and the coupling to the $Z^0$ dominant. In this case, 
the total $e^{*}$ production would be much smaller, and a large 
fraction would have $Q^2>1000$~GeV$^2$, which would be
rejected by the selection criteria. However, even for $f^{\prime
}=-0.75f$,  
35\% of all excited electrons are produced with $Q^2<5$~GeV$^2$.}
The acceptance varies from 40\%
at $M_{e\gamma }=$ 30~GeV to 80\% for $M_{e\gamma }>$ 150~GeV. The
background from neutral current DIS is estimated to be $16.0\pm 2.4$ events,
and the background from QED-Compton scattering is estimated to be $81.5\pm
4.1$ events. The sum agrees well with the 103 events selected from data.

\begin{figure}[p]
\centerline{\epsfbox{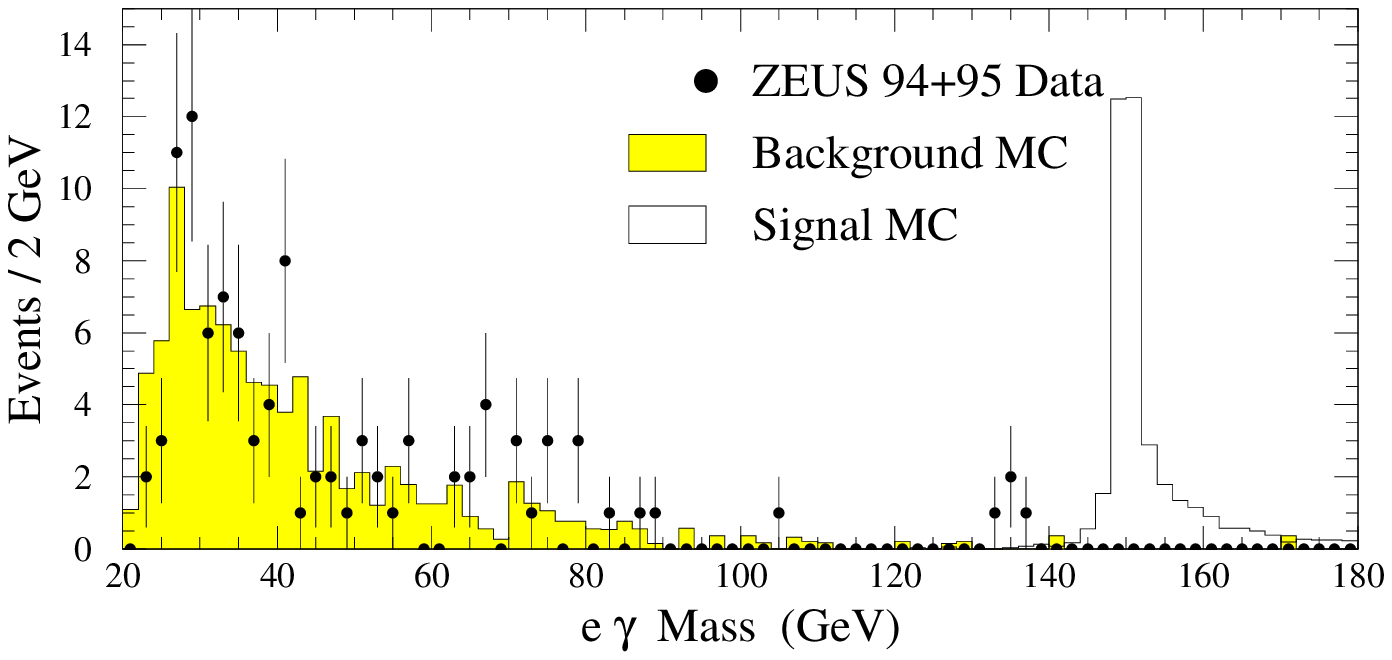}}
\caption{\label{fig:egam_mass}
Distribution of the 
$e\,\gamma$ 
invariant mass for
$e^*\rightarrow e\, \gamma$ 
candidates.  Solid points show ZEUS data, 
and the shaded histogram represents the expected background. 
The unshaded histogram shows the expected lineshape for a 150~GeV
excited electron  (assuming $f=f'$ and 
$f/\Lambda = 8.0\times 10^{-3}$ GeV$^{-1}$).
}
\end{figure}

The electron-photon invariant mass, $M_{e\gamma }$, is determined
using the polar angles, $\theta _e$ and $\theta _\gamma $, of the $e$ and $
\gamma $ clusters, respectively. The formula assumes that the recoiling
hadron system has negligible transverse momentum and that $E-P_{{\rm Z}}=2E_e$
for the $e\,\gamma $ pair: 
\begin{equation}
M_{e\gamma }^2\;=\;(2E_e)^2\,{\frac{\sin \theta _\gamma +\sin \theta _e+\sin
(\theta _\gamma +\theta _e)}{\sin \theta _\gamma +\sin \theta _e-\sin
(\theta _\gamma +\theta _e)}}  \label{massform}
\end{equation}
Here $E_e$ is the electron beam energy. This formula provides a mass
resolution for elastically produced excited electrons nearly a factor of 5
better than that obtained from the measured four-momentum in the calorimeter.
It also reduces the sensitivity to systematic uncertainty in the calorimeter
energy scale. However,
no improvement in the resolution is obtained for inelastically
produced $e^*$ because the recoiling hadron
system has been ignored. 
Non-radiative low $Q^2$ events
tend to occupy the central portion of an $e^{*}$ peak,
while events with initial state radiation would
produce a high 
energy tail in an $e^{*}$ mass peak. 
It follows that the formula (\ref{massform}),
together with the selections, introduces some model dependence in the
acceptance. Nevertheless, low $Q^2$ events are predicted to dominate the
rate for essentially any non-zero coupling to the photon 
(which must be the case for the $e\gamma$ final state to occur).
Hence, if $e^{*}\rightarrow e\,\gamma $ were detected, a narrow central peak
in the $M_{e\gamma }$ distribution derived from (\ref{massform}) is
expected. The unshaded histogram in Fig. \ref{fig:egam_mass} shows the
expected lineshape for an excited electron of mass 150~GeV. The
mass resolution determined from a Gaussian fit to the central region of the
Monte Carlo lineshape varies from 0.5~GeV at a mass of 50~GeV to 1.8~GeV at
a mass of 200~GeV. 

Figure~\ref{fig:egam_mass} shows the invariant mass spectrum
obtained for data
(solid points) compared with the Monte Carlo predicted background
(shaded histogram). 
The Kolmogorov test \cite{Kolmogorov}
applied to the mass distribution, which assigns a probability to the
spectrum {\em shape}, yields a probability of 23\% 
that the observed distribution comes from background. 
Four events are found to cluster at a mass of 135~GeV, which is
significant given the small expected background (2.5 events with 
$M_{e\gamma}>100$~GeV). However, a resonance at $M_{e\gamma}$ should
exhibit a Jacobian peak at $M_{e\gamma}/2$ in the distribution of the
transverse energy of each shower. Instead, three of the four events
have shower transverse energies less than 32~GeV, at the tail of the
expected distribution.  These three events
also exhibit some hadronic energy around the photon candidate, and the
EM clusters are near the minimum energy or angle requirements.  
The features of these three events are consistent with neutral current
DIS background.
The fourth event, with an average shower transverse energy of 54~GeV,
 has no hadronic energy, but both EM clusters have matching tracks. 
It is consistent with an event in which the photon 
converts near the vertex. 
We conclude that the data are consistent with background
expectations.

\subsection{$e\,q\,\overline{q}$ Resonance Search}

\label{sec:ezhad}

The $e\,q\,\overline{q}$ final state could arise from $e^{*}\rightarrow
e\,Z^0$ or $\nu _{{\rm e}}^{*}\rightarrow e\,W$ decays in which the $Z^0$ or 
$W$ decays hadronically (additional $Z^0$ and $W$ decay modes are described
in Sections~\ref{sec:enunu} and \ref{sec:eenu}). The search
makes use of (a) selections to obtain events with 
the characteristics expected for this
decay mode; and (b) calculation of the final state mass ($M_{eZ}$ or
$M_{eW}$).
No distinction is made between
excited electrons and excited neutrinos for purposes of optimizing cuts
and calculating backgrounds in this search, since both involve a
topology with a high 
energy electron and two jets. With $\nu _{{\rm e}}^{*}$ production
proceeding through $W$ exchange, the characteristic $Q^2$ of the exchanged
boson is typically much larger than for $e^{*}$ production. The only
difference in the final state topology aside from the
$q\,\overline{q}$ mass, therefore, involves the elasticity
of the final state hadronic system associated with the struck proton. The
dominant background to this search arises from multi-jet neutral current
DIS. Global event variables (without explicit jet algorithms) are used to
separate the signal from this background.

For this search, an EM cluster with transverse energy larger than $15$~GeV
must be identified with $\theta <115^{\circ }$. The shower isolation 
requirement (see Section~\ref{sec:kinematics}) is dropped
because the electron from an $f^{*}$ decay is often near some hadronic
activity when produced near threshold. 
The measured longitudinal momentum variable, $E-P_{{\rm Z}}$, must be in
the range 35 to 65~GeV. Large hadronic transverse energy ($E_{{\rm t,had}}$)
and large invariant mass ($M_{{\rm had}}$), discussed in Section~\ref
{sec:kinematics}, must be measured from the calorimeter. Specifically,
either $E_{{\rm t,had}}>60$~GeV and $M_{{\rm had}}>60$~GeV, or $E_{{\rm t,had
}}>70$~GeV and $M_{{\rm had}}>40$~GeV is required. The latter selection
provides good acceptance for high mass $f^{*}$ decays in which hadrons from
the decay of the heavy boson are often lost in the very forward region.
Remaining cosmic ray backgrounds are suppressed by requiring approximate
momentum balance in the transverse plane: {\mbox{$\not\hspace{-.55ex}{P}_{
\rm t}$}}/$E_{{\rm t}}$$<0.15$.

\begin{figure}[p]
\centerline{\epsfbox{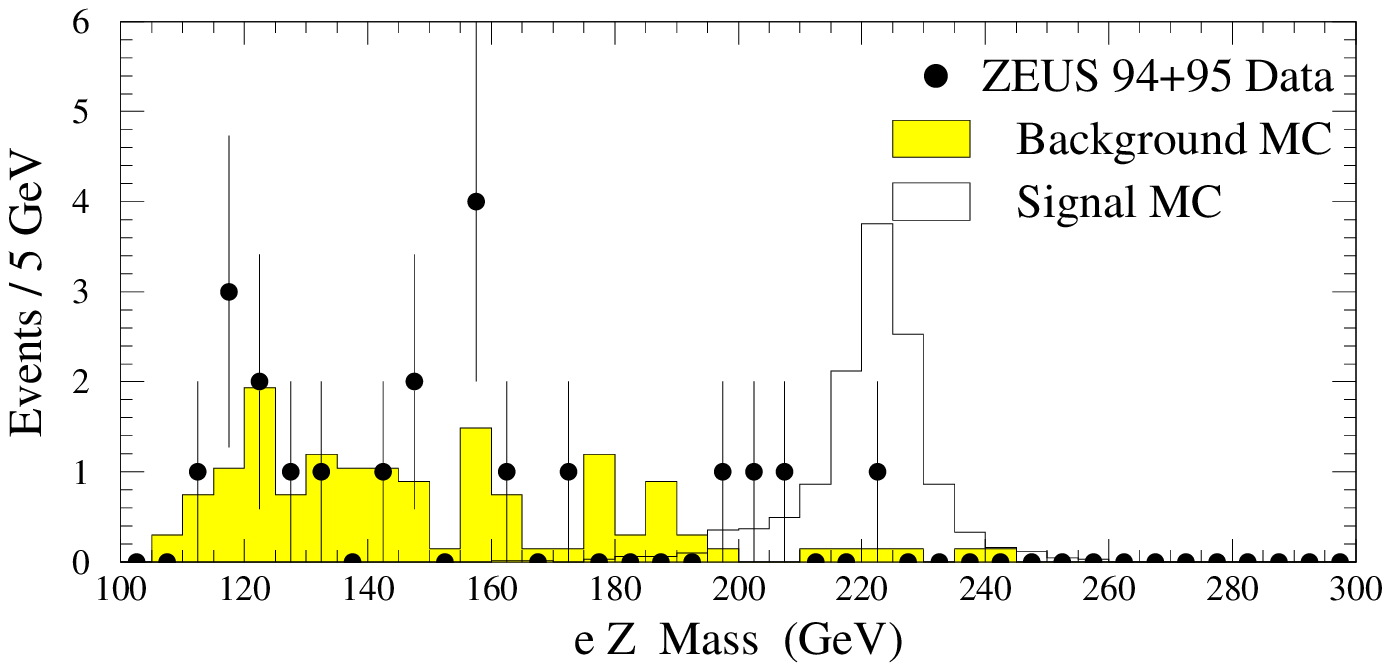}}
\caption{\label{fig:ez_mass}
Distribution of the $e\, Z$ invariant mass for 
$e^*\rightarrow e\, Z^0 \rightarrow e\,q\,\overline{q}$ candidates.
Solid points show ZEUS data, 
and the shaded histogram represents the expected background. 
The unshaded histogram shows the expected lineshape for a 225~GeV
excited electron (assuming $f=f'$ and 
$f/\Lambda = 5.0\times 10^{-2}$ GeV$^{-1}$).
}
\end{figure}

The electron-boson invariant mass, $M_{eV}$, is calculated from the energy
and angle of the decay electron because it is 
more accurate than
the invariant mass calculated from the entire calorimeter, which
suffers from leakage in the very forward region and the
poorer hadronic energy resolution.
The formula, derived assuming the $f^{*} 
$ has negligible transverse momentum and $E-P_{{\rm Z}
}=2E_e$, is
\begin{equation}
M_{eV}^2\;=\;{\frac{4E_eE_e^{\prime }\cos ^2{\frac{\theta _e}2}+M_V^2}{
1-E_e^{\prime }/E_e\sin ^2{\frac{\theta _e}2}}}  \label{eq:ezmass}
\end{equation}
where $E_e$ is the electron beam energy, $E_e^{\prime }$ is the decay
electron energy, $\theta _e$ is the polar angle of the decay electron, and $
M_V$ is the mass of the gauge boson. The mass resolution, determined from
Monte Carlo simulations, is approximately 6~GeV for excited
electrons and 18~GeV for excited neutrinos, the latter being larger
because the underlying assumptions of Eq.~(\ref{eq:ezmass}) are not
rigorously correct when $Q^2$ is large.
 The unshaded histogram in Fig.~\ref
{fig:ez_mass} shows the expected lineshape for an $e^{*}$ of mass 225~GeV.

The ratio of the total calorimeter invariant mass (Eq.~\ref{eq:sumhad}), 
$M$, to the mass derived in Eq.~(\ref{eq:ezmass}) is used 
to further reject
backgrounds.  Approximately one-third
of the remaining background is removed when the cut $M/M_{eZ}>0.82$ is
applied.

Above threshold the acceptance for the $e\,q\,\overline{q}$ final state is
approximately 55\% for $e^{*}$ production, and 70\% for $\nu _{{\rm e}}^{*}$
production. The total number of selected events is 21, compared with a
background expectation of $15.3\pm 1.5$. The distribution in $M_{eZ}$ for
the selected events with the $e^{*}\rightarrow e\,Z^0$ hypothesis is shown
in Fig.~\ref{fig:ez_mass} for data (solid points) compared with the Monte
Carlo background (shaded histogram). No significant excess is observed. The
Kolmogorov test applied to the 
mass distribution yields a
probability of 86\% that the shape of the observed distribution is
consistent with the expected background.

\subsection{$\nu_{{\rm e}}\,q\,\overline{q}$ Resonance Search}

\label{sec:nuwhad}

The $\nu _{{\rm e}}\,q\,\overline{q}$ final state would arise from $
e^{*}\rightarrow \nu _{{\rm e}}\,W$ or $\nu ^{*}\rightarrow \nu _{{\rm e}
}\,Z^0$ decays where the $Z^0$ or $W$ decays hadronically (an additional $W$
decay mode is $e^{*}\rightarrow \nu _{{\rm e}}\,W\rightarrow e\,\nu _{{\rm e}
}\,\overline{\nu }_{{\rm e}}$, described in Section~\ref{sec:enunu}). The
technique, using event topology and invariant mass, 
is similar to that in Section \ref{sec:ezhad}.  Again 
the same selection is used for
the excited electron and excited neutrino 
searches.  The
dominant background consists of multi-jet charged current
DIS events.

The presence of a neutrino is inferred from substantial missing 
momentum measured in the calorimeter: {\mbox{$\not\hspace{-.55ex}{P}_{\rm t}$}}
$>15$~GeV
and $10<E-P_{{\rm Z}}<45$~GeV. 
To further suppress non-$e\,p$ collisions
we also require {\mbox{$\not\hspace{-.55ex}{P}_{\rm t}$}}$>$
10~GeV for calorimeter cells with $\theta >10^{\circ }$.
The hadron system
must exhibit large transverse energy ($E_{{\rm t}}>40$~GeV), and large
invariant mass ($M>55$~GeV). The polar angle of the final state neutrino,
calculated from the hadron measurements assuming that the 
$\nu _{{\rm e}}\,q\,\overline{q}$ system conserves the
longitudinal momentum variable ($E-P_{{\rm Z}}=2E_e$) and transverse
momentum ({\mbox{$\not\hspace{-.55ex}{P}_{\rm t}$}}$=0$), 
is required to be less than $140^{\circ }$.

The acceptance for the $\nu _{{\rm e}}\,q\,\overline{q}$ final state
is approximately 65\% for $e^{*}$ production, and 75\% for $\nu ^{*}$
production. The 
total number of events observed in data is 13, compared with a background
expectation of $8.4\pm 1.5$ predominantly from charged current DIS, as seen
in Tab.~\ref{tab:effic}.

\begin{figure}[p]
\centerline{\epsfbox{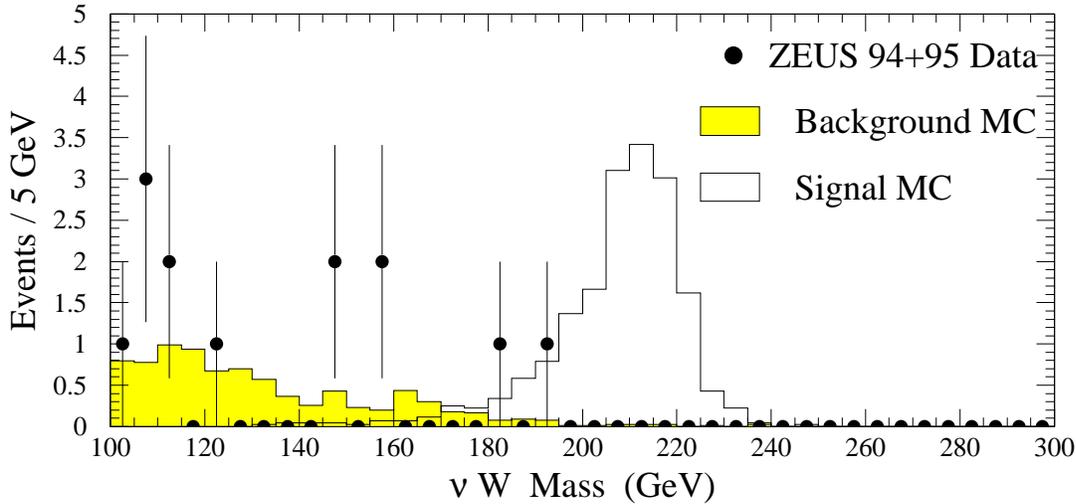}}
\caption{\label{fig:nuw_mass}
Distribution of the $\nu_{\rm e}\, W$ invariant mass for 
$e^*\rightarrow \nu_{\rm e}\, W \rightarrow \nu_{\rm e}\,q\,\overline{q}$ 
candidates.
Solid points show ZEUS data, 
and the shaded histogram represents the expected background. 
The unshaded histogram shows the expected lineshape for a 225~GeV
excited electron (assuming $f=f'$ and 
$f/\Lambda = 2.5\times 10^{-2}$ GeV$^{-1}$).
The peak is shifted to lower mass because the measured hadronic energy
is not explicitly corrected for losses.
}
\end{figure}

The mass of the neutrino-boson system, $M_{\nu V}$, 
is calculated from the measured quantities {
\mbox{$\not\hspace{-.55ex}{P}_{\rm t}$}}\ and $\delta \equiv E-P_{{\rm Z}}$.
With the assumptions described above, the invariant mass is 
\begin{equation}
M^2_{\nu V} \; =\; {\frac{4\not{\hspace{-.55ex}}{P}_{{\rm t}}^2E_e^2
+2E_eM_V^2(2E_e-\delta) }{\delta(2E_e-\delta) }}  \label{eq:nuwmass}
\end{equation}
Here $E_e$ is the electron beam energy and $M_V$ is the mass of the gauge
boson.  The unshaded histogram in Fig.~\ref{fig:nuw_mass} 
shows the expected lineshape for an excited electron of mass 225~GeV.  
The mass resolution determined from Monte Carlo simulations is
approximately 9~GeV (14~GeV) for excited electrons (excited neutrinos).

The invariant mass distribution for the selected events with 
the $e^*\rightarrow \nu_{{\rm e}}\, W$ hypothesis is
shown in Fig.~\ref{fig:nuw_mass} for data (solid points) compared with the
Monte Carlo background (shaded histogram). No evidence for a peak is seen.
The Kolmogorov test applied to the 
mass distribution yields a probability of 55\% 
that the shape of the observed distribution arises
from the expected background.

\subsection{$\nu_{{\rm e}}\, \gamma$ Resonance Search}

This final state could occur if neutrinos contained charged
constituents. The principal signature is an isolated electromagnetic cluster
in events with missing transverse momentum. Backgrounds arise from charged
current DIS events with isolated $\pi ^0$s or initial state radiation.

A photon is tagged by identifying an EM cluster with more than $15$~GeV of
transverse energy. The cluster is required to be isolated in an 
$\eta$--$\phi$ 
cone of radius $R=0.7$ (see Section~\ref{sec:kinematics}). A photon
candidate is rejected if a track measured by the central tracking detector
with $\theta >15^{\circ }$ projects to within 40~cm of the cluster.

The presence of a neutrino is inferred by requiring {\mbox{$\not
\hspace{-.55ex}{P}_{\rm t}$}}$>15$~GeV. The recoil jet typically associated
with $\nu ^{*}$ production is identified by requiring $E_{{\rm t,had}}>10$
~GeV. Neutral current DIS events are suppressed by requiring $E-P_{{\rm Z}
}<50$~GeV.


No event survives the selection criteria.
The background expected from charged
current DIS is $0.4\pm 0.1$ events. The acceptance for the $\nu _{
{\rm e}}\,\gamma $ final state is approximately 50\%.

\subsection{$e\,\nu \,\overline{\nu }$ Resonance Search}

\label{sec:enunu}

This final state could arise from production of an $e^{*}$, with subsequent
decays: $e^{*}\rightarrow e\,Z^0$ and $Z^0\rightarrow \nu \overline{
\nu }$; or $e^{*}\rightarrow \nu _{{\rm e}}\,W$ and $W\rightarrow e\,\nu _{
{\rm e}}$.\footnote{
Other decays which lead to missing transverse momentum in the
calorimeter (e.g., $Z^0\rightarrow \mu ^{+}\mu ^{-},\tau ^{+}\tau ^{-}$, etc.)
would also be selected with lower efficiency; these decays are not used 
in calculating the sensitivity.} 
Such events would contain a striking signature:
one high $E_{{\rm t}}$ electron in the detector, and nothing else except for
a possible low $E_{{\rm t}}$ recoil jet. Backgrounds from charged and
neutral current DIS are small. A rare Standard Model background
process is $W$ production.

The electron is tagged by identifying an EM cluster with more than $15$~GeV
of transverse energy. The cluster must be isolated in an $\eta $--$
\phi $ cone of radius $R= 0.7$ and be accompanied
by a track with momentum
larger than 10~GeV that projects to within 40~cm of the cluster.

Events are expected to have missing transverse and longitudinal
momentum, so {\mbox{$\not\hspace{-.55ex}{P}_{\rm t}$}}$>15$~GeV 
and $E-P_{\rm Z}<45$~GeV are required.
If a recoil jet is present, it
is not expected to be back-to-back in azimuth with the electron
because of the transverse momentum carried by the final state neutrinos.
Consequently, if the hadronic transverse energy $E_{{\rm t,had}}$ is larger
than $2$~GeV, we require $\cos (\phi _{{\rm e}}-\phi _{{\rm had}})>-0.95$.
The acceptance for this final state is approximately 70\% for $
e^{*}\rightarrow e\,Z^0$ decays, and 60\% for $e^{*}\rightarrow \nu _{{\rm e}
}\,W$ decays.

One event survives the selection criteria, with mass
$M_{eZ} = 116$~GeV.
The expected background, itemized in Tab.~\ref{tab:effic}, is $1.0\pm 0.2$
events.

\subsection{$e\,\overline{e}\,\nu_{{\rm e}}$ Resonance Search}

\label{sec:eenu}

The selection for $\nu _{{\rm e}}^{*}\rightarrow e\,W\rightarrow 
e\,\overline{e}\,\nu _{
{\rm e}}$ is a variation of the $e\,\nu \,\overline{\nu }$ selection
criteria discussed in the previous section. These decays would feature two
high transverse energy EM clusters, missing transverse momentum, and a
possible recoil jet from the struck proton.

We require two EM clusters, each with more than $
10$~GeV of transverse energy. No isolation cut is applied, and no explicit
track match is made. The presence of a neutrino is inferred by requiring {
\mbox{$\not\hspace{-.55ex}{P}_{\rm t}$}}$>15$~GeV. Since the two electrons
can carry a substantial amount of the $\nu _{{\rm e}}^{*}$ momentum, only
the loose cut $E-P_{{\rm Z}}<65$~GeV is applied.


The acceptance for the 
$e\, \overline{e}\, \nu_{{\rm e}}$
final state is approximately 60\%.
No event survives the cuts. The expected background is very
small, and is assumed negligible for the purposes of setting cross
section upper limits. 

\subsection{$q\,\gamma$ Resonance Search}

\label{sec:qgam}

Events with excited quarks ($q^{*}$) share some characteristics with low $
Q^2 $ DIS events and with photoproduction processes, in that the scattered 
electron might enter the detector or travel in the
direction of the electron beam and miss the detector. Hence, one cannot cut
tightly on $E-P_{{\rm Z}}$, and the cut
on $E_{{\rm rear}}$ must be relaxed from that used for the excited lepton
searches. We require either $E_{{\rm rear}}<2$~GeV, or else $E_{{\rm 
rear}}<10$~GeV and the electromagnetic fraction of the
calorimeter energy in 
the rear direction is larger than 90\%. This removes most low~$Q^2$ neutral
current DIS events and resolved photoproduction events, while retaining
efficiency for excited quarks which typically have only a low energy
electron in the rear direction.

A resonance in the photon-jet invariant mass spectrum would provide
compelling evidence for the decay $q^*\rightarrow q\,\gamma$. Photons are
selected by identifying an EM cluster with 
more than $15$~GeV of transverse
energy. The cluster is required to be isolated in a large $\eta $--$\phi $
cone of radius $R=1.5$, a larger cone than used in the other searches to
reduce the large rate from $\pi ^0$s produced in the jets of
photoproduction reactions. A photon candidate is rejected if a track
measured by the central tracking detector with $\theta >15^{\circ }$
projects to within 40~cm of the EM cluster.

The photon and hadron (see Sec. \ref{sec:kinematics}) polar angles must be in
the forward region: $\theta _\gamma <80^{\circ }$and $\theta _{{\rm had}
}<100^{\circ }$. The total transverse energy measured for the event must be
large: $E_{{\rm t}}$$>30$~GeV.

\begin{figure}[p]
\centerline{\epsfbox{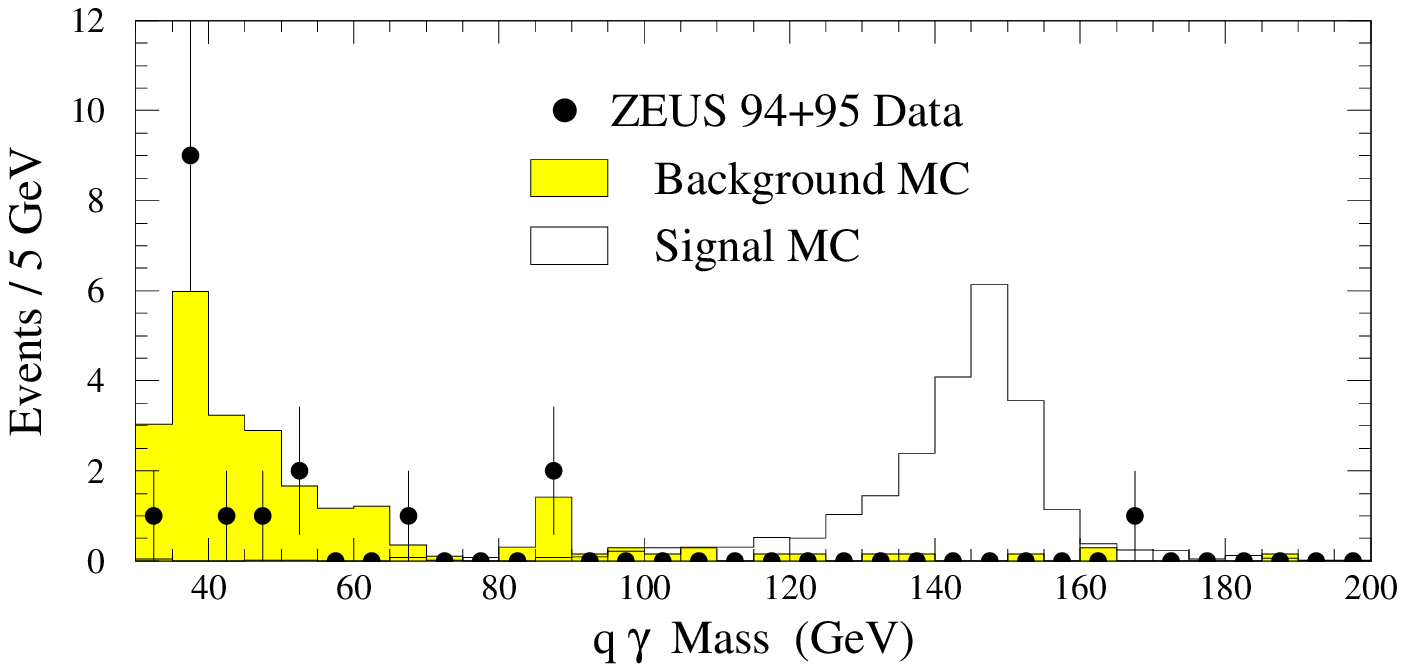}}
\caption{\label{fig:qgam_mass}
Distribution of the 
$q\,\gamma$ 
invariant mass for
$q^*\rightarrow q\, \gamma$ 
candidates.  Solid points show ZEUS data, 
and the shaded histogram represents the expected background. 
The unshaded histogram shows the expected lineshape for a 150~GeV
excited quark (assuming $f_{\rm s} = 0$, $f=f'$, and 
$f/\Lambda = 1.0\times 10^{-2}$ GeV$^{-1}$).
}
\end{figure}

The measured longitudinal momentum variable will not in general satisfy the
relation, $E-P_{{\rm Z}}=2E_e$, for events containing excited quarks. 
However, the heavy $q^{*}$ is produced with little transverse momentum,
which permits the $q\gamma $ invariant mass, $M_{q \gamma}$,  
to be calculated optimally from $E_{\gamma}$, $\theta_{\gamma}$, 
and $\theta _{{\rm had}}$:
\begin{equation}
M_{q\gamma}^2 \; = \; 2\,E_{\gamma}^2 \,
{\sin\theta_\gamma \over \sin\theta_{\rm had}} \,
\Big[ 1 - \cos(\theta_\gamma+\theta_{\rm had}) \Big]
\label{eq:qgmass}
\end{equation}
The mass resolution determined
from Monte Carlo simulations varies from 3~GeV at a mass of 40~GeV to 8~GeV
at a mass of 150~GeV. The unshaded histogram in Fig.~\ref{fig:qgam_mass}
shows the expected lineshape for an excited quark of mass 150~GeV.

The overall acceptance for events with $q^{*}\rightarrow q\,\gamma $ varies
from 20\% at a mass of 40~GeV to 65\% for masses above 150~GeV. The total
number of events selected from data is 18, compared with a total background
expectation of $23.5\pm 2.5$ events, tabulated in detail in Tab. \ref
{tab:effic}.

The $q\,\gamma $ invariant mass distribution for the selected events is
presented in Fig.~\ref{fig:qgam_mass} for data (solid points) compared with
the Monte Carlo background (shaded histogram). The spectrum shows no
evidence for a resonance. The Kolmogorov test applied to the 
mass spectrum yields a probability of 47\% that the 
observed distribution
conforms to the expected background shape. 

\subsection{$q\,\overline{e}\,\nu _{{\rm e}}$ Resonance Search}

We address here the case of excited quarks decaying through the chain: $
q^{*}\rightarrow q\,W$, with $W\rightarrow \overline{e}\,\nu_{\rm e}$.
The final state contains 
a neutrino, an electron, and
a jet which is not back-to-back in azimuth with the electron.
Each has high transverse momentum, 
so backgrounds from DIS and
photoproduction are expected to be minimal. The selection criteria for the $
q\,\overline{e}\,\nu _{{\rm e}}$ search are similar to those used for 
the $e\,\nu 
\,\overline{\nu }$ search, except that the hadron system
from the decay quark must be identified and the cut on $E_{{\rm rear}}$ is
as discussed in the previous section.

The electron is tagged by identifying an EM cluster with more than $15$~GeV
of transverse energy. The cluster is required to be isolated in an 
$\eta $--$\phi $ cone of radius $R=0.7$. No explicit track match is demanded.
Since a high transverse momentum neutrino is expected, {\mbox{$\not
\hspace{-.55ex}{P}_{\rm t}$}}$>15$~GeV is imposed. The upper bound on $E-P_{
{\rm Z}}$ is relaxed to 65~GeV since the scattered electron may enter the
detector. The energetic hadron system is identified 
by selecting events with $E_{{\rm 
t,had}}>10$~GeV. Since the hadronic jet is not expected to be back-to-back in
azimuth with the electron, the cut $\cos (\phi _{{\rm e}}-\phi _{{\rm had}
})>-0.95$ is applied.

The acceptance for the $q\,\overline{e}\,\nu _{{\rm e}}$ final state is
approximately 50\%. No event survives the selection criteria.
The total background expected is $1.5\pm 0.3$ events, roughly equally from
neutral current and charged current DIS processes, as described
in Tab. \ref{tab:effic}.

\section{Systematic Uncertainties}

\label{sec:systematics}

The overall uncertainty on the normalizations for the backgrounds and
signals in these searches receives
contributions from the acceptance and luminosity, as
well as theoretical uncertainties on the production mechanisms. Our
evaluation of these are itemized below. Experimental errors are listed
first.

\begin{itemize}
\item  The uncertainty on the measured integrated luminosity of the combined
1994 and 1995 $e^{+}p$ data sample is 1.5\%.

\item  The uncertainty on the electron/photon identification
efficiency is estimated to be
at most 5\%, determined by comparing alternate algorithms for EM
cluster identification.

\item  The sensitivity of event selection to vertex reconstruction algorithms 
and the underlying vertex distribution is estimated to produce uncertainty
at the 4\% level.

\item  The systematic 3\% uncertainty in the calorimeter energy scale is
found to lead to an overall 3\% uncertainty in the acceptance.

\item  Since calculations use smooth parameterizations of the excited
fermion acceptances as functions of the $f^*$ mass, some error is
incurred by interpolation between 
generated Monte Carlo mass points. This uncertainty is estimated to be 4\%.

\item  The effects of errors on background estimates used in the upper limit
derivation depend on the number of
observed events as well as the size of the background. However, 10\%
variations of the background normalization typically lead to variations of
6\% or less on upper limits.

\item  The theoretical uncertainty introduced by radiative corrections to
the excited fermion production model and the uncertainty associated with the
parton density distributions used to model the proton is taken to be
8\%, as determined from our earlier study \cite{ZEUS}.

\item  The model dependence of the excited fermion angular decay
distribution is evaluated by comparing the acceptance for isotropic 
decays versus the nominal \cite{Hagiwara,Boudjema} distribution.\footnote{%
For $f^{*} \rightarrow f \gamma$, for example, the nominal decay distribution 
is ($1+\cos \theta ^{*}$), where $\theta ^{*}$ is the polar angle 
between the incoming and outgoing fermion in the $f^{*}$ rest frame.}
The deviation is typically 5\% or less.
\end{itemize}

Adding all contributions in quadrature yields
a total systematic uncertainty of 14\%, which is incorporated into the upper
limit procedure described in the Appendix.

\section{Results}

\label{sec:results}

We have no positive evidence for excited leptons or quarks in any of
the 8 searches 
(11 excited fermion decay chains) described above. The data provide 
upper limits (U.L.) as a function of the excited fermion mass, $M_{f^{*}}$,
on the production cross section times branching ratio: 
\begin{equation}
(\sigma \cdot BR)_{{\rm U.L.}}\,=\,{\frac{N_{{\rm U.L.}}\,(M_{f^{*}})}{
A(M_{f^{*}})\cdot L}}
\end{equation}
where $L$ is the integrated luminosity and $A(M_{f^{*}})$
is the parameterized acceptance for the excited fermion decay
including, when appropriate, the branching fraction of the $Z^0$ or
$W$ decay. 
The upper limit on the number of signal events, $N_{{\rm 
U.L.}}$, is calculated at the 95\% confidence level as a function of mass
according to an unbinned likelihood technique. This procedure, including
its use to obtain combined upper limits for cases with more than
one decay chain, is described in the Appendix.

\begin{figure}[p]
\centerline{\epsfbox{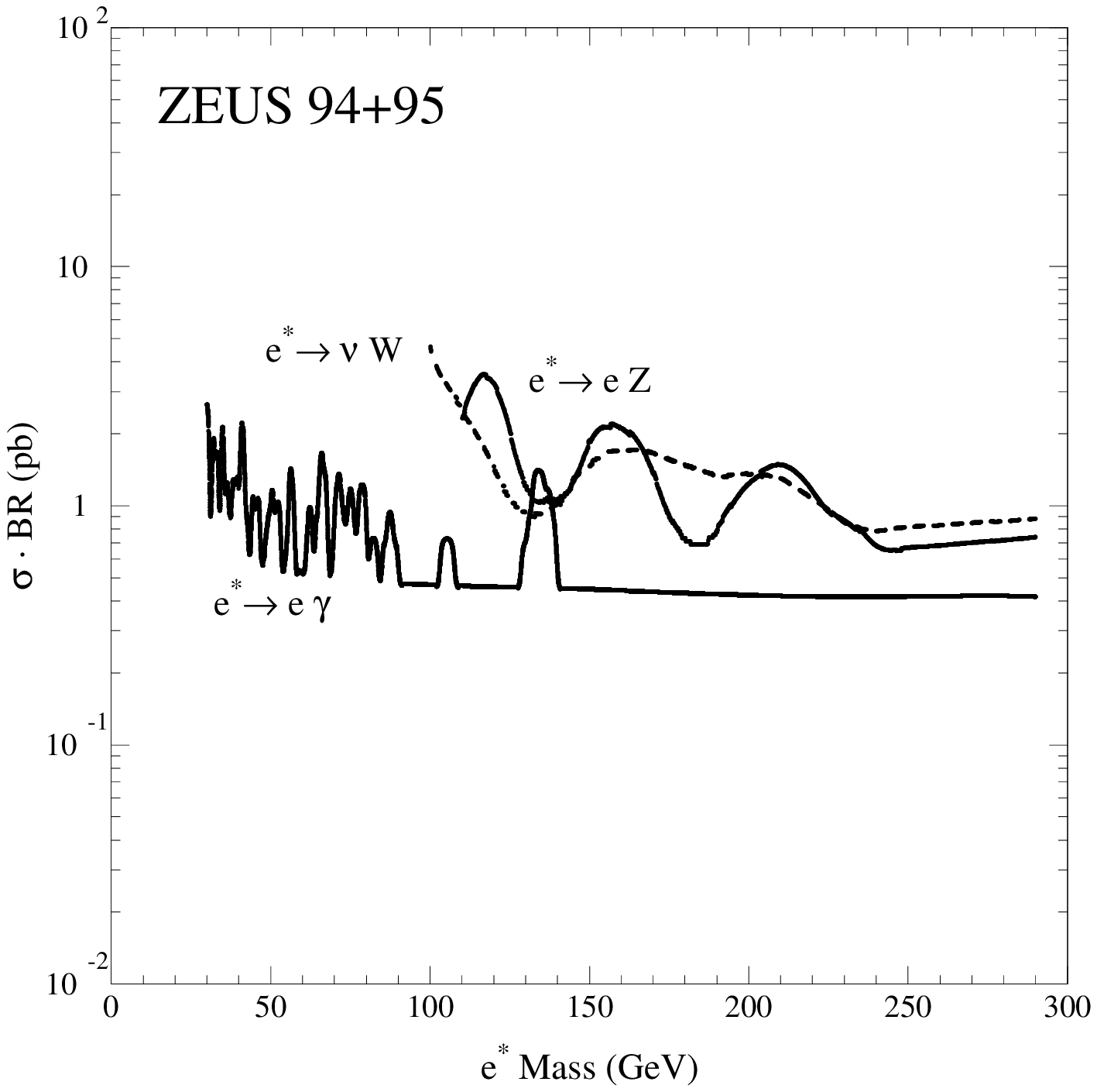}}
\caption{\label{fig:estar_sigma}
Upper limits at the 95\% confidence level on the product of
the production cross section (in picobarns) for the process
$e^+\,p\rightarrow e^*\,X$ 
and the branching ratio for $e^*\rightarrow \ell\, V$
as a function of mass for the studied $e^*$ channels.
}
\end{figure}

\begin{figure}[p]
\centerline{\epsfbox{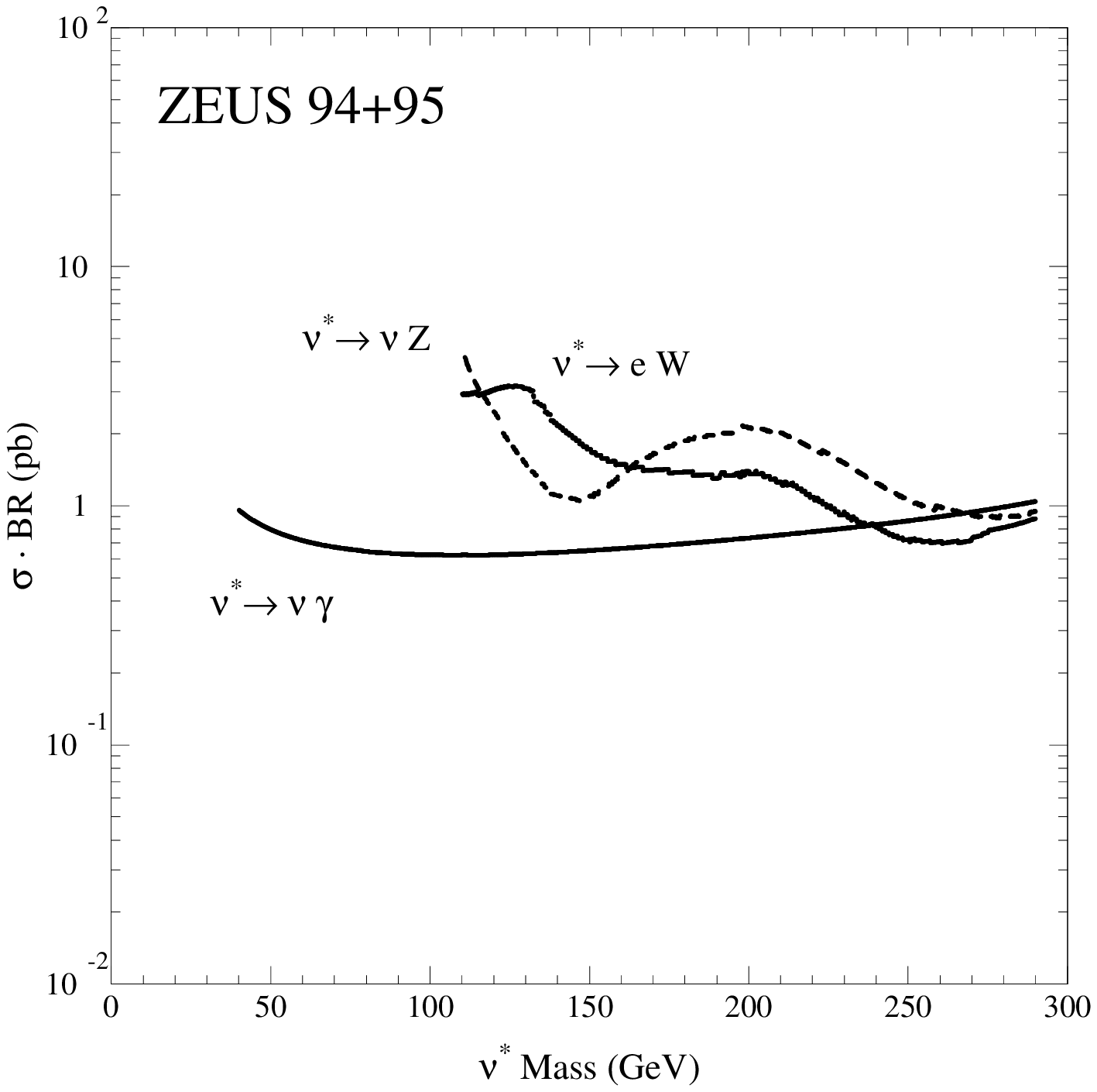}}
\caption{\label{fig:nstar_sigma}
Upper limits at the 95\% confidence level on the product of
the production cross section (in picobarns) for the process
$e^+\,p\rightarrow \nu^*_{\rm e}\,X$ 
and the branching ratio for $\nu^*_{\rm e}\rightarrow \ell\, V$
as a function of mass for the studied $\nu_{\rm e}^*$ channels.
}
\end{figure}

\begin{figure}[p]
\centerline{\epsfbox{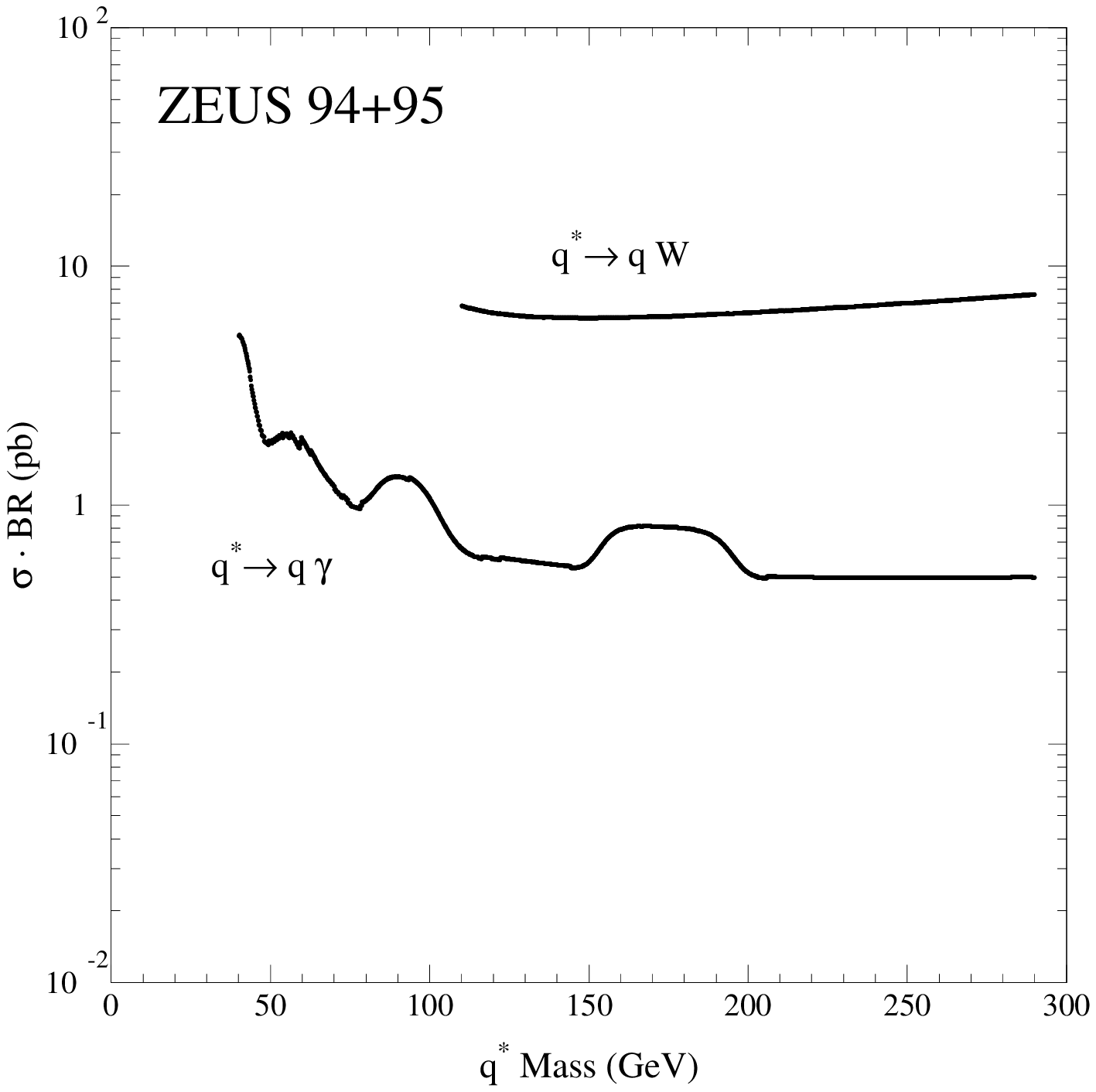}}
\caption{\label{fig:qstar_sigma}
Upper limits at the 95\% confidence level on the product of
the production cross section (in picobarns) for the process
$e^+\,p\rightarrow e^+\,q^*\,X$ 
and the branching ratio for $q^*\rightarrow q\, V$
as a function of mass for the studied $q^*$ channels.
}
\end{figure}

The resulting upper limit curves on the production cross sections are shown
in Fig.~\ref{fig:estar_sigma} for excited electrons, Fig.~\ref
{fig:nstar_sigma} for excited neutrinos, and Fig.~\ref{fig:qstar_sigma} for
excited quarks.
For final states with photons, the
limits are typically less than 1~pb, an order of
magnitude improvement over our previous upper limits \cite{ZEUS}. 
The limits are below those reported by H1 \cite{H1} 
because this ZEUS sample uses a larger integrated luminosity.

We also set upper limits on coupling strengths in the compositeness model
described by the Lagrangian~(\ref{eq:Baur}), with the relationship fixed
among the $ 
SU(2)_L$, $U(1)_Y$, and $SU(3)_C$ coupling constants: $f$, $
f^{\prime }$, and $f_{{\rm s}}$, respectively.
Branching fractions are then determined
and cross sections calculated in terms of a single
unknown parameter,  $f/\Lambda $. The upper limit at the
95\% confidence level is obtained from the upper limit 
$(\sigma \cdot BR)_{{\rm U.L.}}$
by using the relation 
\begin{equation}
\Bigg({\frac f\Lambda }\Bigg)_{{\rm U.L.}}\,=\,\sqrt{\frac{(\sigma \cdot
BR)_{{\rm U.L.}}}{\sigma _{{\rm MC}}\cdot BR}}\;\Bigg({\frac f\Lambda }\Bigg)
_{{\rm MC}}
\end{equation}
where $BR$ is the branching ratio, and $\sigma _{{\rm MC}}$ is the
Monte Carlo prediction for the 
theoretical cross section  calculated using the
coupling  $({f/\Lambda})_{{\rm MC}}$. 

For excited leptons, the parameter $f_{{\rm s}}$ is irrelevant.  
For $e^{*}$, we choose $f=f^{\prime }$.  For $\nu_{\rm e}^{*}$,  
we set  $f=-f^{\prime }$ 
(see Section \ref{sec:theory}).

For excited quarks, we choose $f=f^{\prime }$ and $f_{{\rm s}}=0$.
Given the stringent limits\cite{CDFold,CDFnew} 
on $f_{\rm s}$ set by CDF on $q^*$ production,
we have chosen to concentrate on
the unique HERA sensitivity at high mass 
to electroweak couplings: $f$ and $f^{\prime }$.  
We further assume transitions to a single excited mass degenerate doublet 
$(u^{*},d^{*})$ such that the production cross section is the sum of
both $u$ and $d$ 
quark excitations.\footnote{
The branching ratio used in the calculation is a weighted average; the
weights are based on the square of the quark charges and the $x$-dependent
relative density of $u$ and $d$ quarks in the proton. Because $
q^{*}$ are produced via photon exchange at HERA, our limits are primarily
sensitive to $u$ quark excitations.}

\begin{figure}[p]
\centerline{\epsfbox{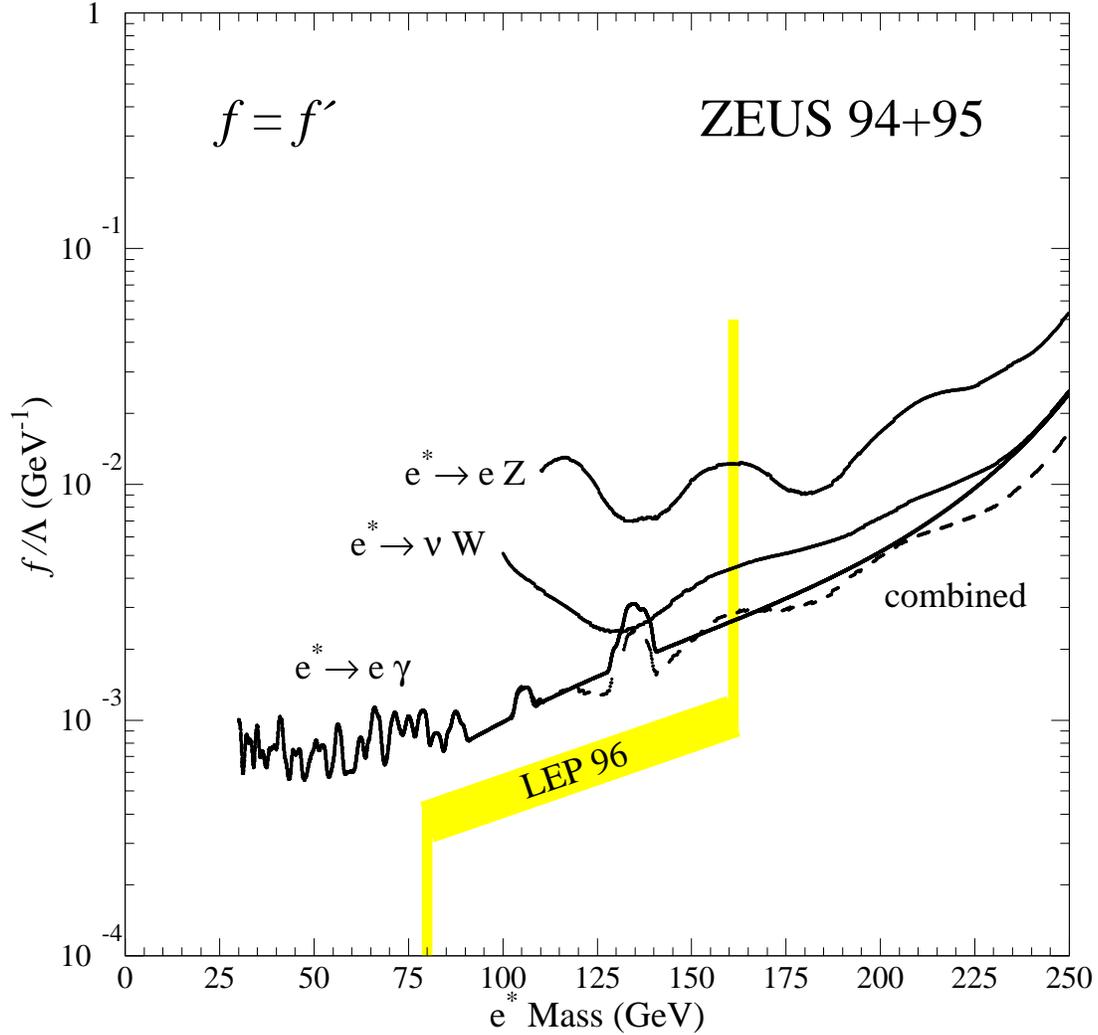}}
\caption{\label{fig:estar_flam}
Upper limits at the 95\% confidence level on 
the coupling $f/\Lambda$ 
as a function of mass for the excited electron channels assuming
$f=f'$.  The dashed line is the combined limit from all three decay modes.
The shaded line is the limit reported recently by the
LEP experiments \protect\cite{LEPonefive,LEPtwo}, where the limit
variations between 80 and 161~GeV are given approximately by the line
width. 
}
\end{figure}

\begin{figure}[p]
\centerline{\epsfbox{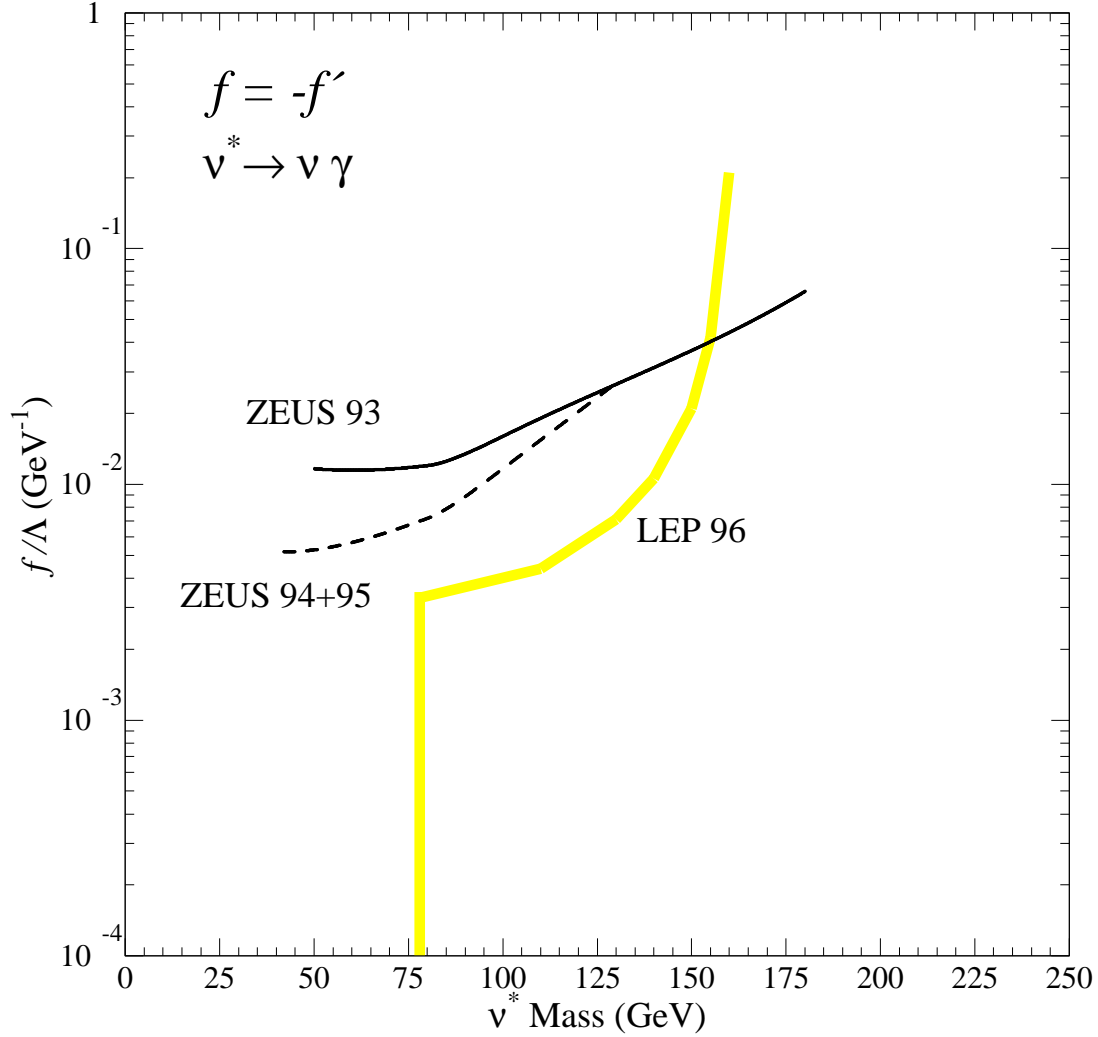}}
\caption{\label{fig:nstar_flam}
Upper limits at the 95\% confidence level on 
the coupling $f/\Lambda$ 
as a function of mass for the
$\nu^*_{\rm e}\rightarrow \nu_{\rm e}\,\gamma$ channel only assuming
$f=-f'$. The dashed line is the limit derived from this analysis,
whereas the solid line is the corresponding limit from our previous
analysis \protect\cite{ZEUS} of $e^-p$ collisions.
The limit from the LEP experiments \protect\cite{LEPonefive,LEPtwo} 
is shown by the shaded line.
}
\end{figure}

\begin{figure}[p]
\centerline{\epsfbox{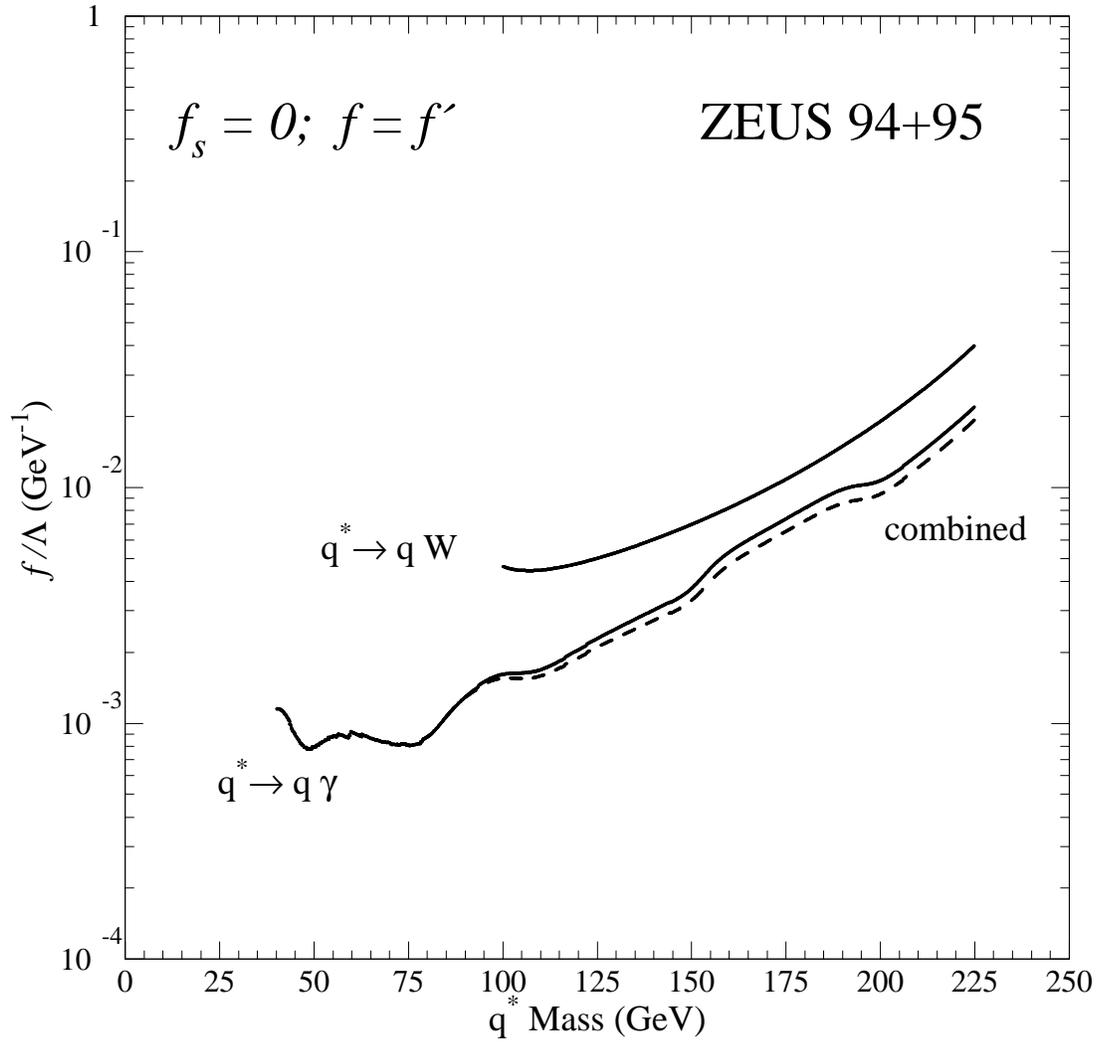}}
\caption{\label{fig:qstar_flam}
Upper limits at the 95\% confidence level on 
the coupling $f/\Lambda$
as a function of mass for the excited quark channels assuming
$f_{\rm s}=0$ (no strong coupling) and $f=f'$.  The dashed line is the
combined limit from both decay modes. 
}
\end{figure}

The upper limits on $f/\Lambda $ as a function of mass
are shown in Fig.~\ref{fig:estar_flam} for $e^{*}$, in
Fig.~\ref{fig:nstar_flam} for $\nu ^{*}_{\rm e}$, 
and in Fig.~\ref{fig:qstar_flam}
for $q^{*}$.  These limits are derived under the narrow width
approximation. For higher masses and couplings than reported here,
the natural width becomes much larger than the experimental width.
The $e^{*}$ and $q^{*}$
searches encompass multiple decay modes; the combined limits are shown 
as dashed lines in each figure. Figure~\ref{fig:nstar_flam} shows the limit
on  $\nu _{{\rm e}}^{*}$ from the previous search by ZEUS \cite{ZEUS} 
in $e^-{p}$ collisions with a
much lower luminosity (0.55 pb$^{-1}$). This previous
limit is superior to that presented here for $M_{\nu ^{*}}>130$~GeV because
the $e^{-}p$ cross section is significantly higher at large
$\nu_{\rm e}^{*}$ masses.\footnote{
The cross section for massive $\nu^*$ production in $e^{+}p$
scattering is heavily suppressed relative to $e^{-}p$ scattering,
partly because of the smaller density of valence $d$ quarks to
valence $u$ quarks (large Bjorken $x$ is required), 
but mainly because the chiral
nature of the $W$ exchange 
results in suppressed particle-antiparticle coupling amplitudes at
large energy transfers.
For $M_{\nu ^{*}}>130$ GeV, the cross
section advantage for $e^{-}$ beam exceeds the luminosity advantage for $
e^{+}$ beam. High luminosity with $e^{-}$ beam will substantially improve
these limits.}

The coupling limit from the excited electron search corresponds to a
lower limit on the compositeness scale $\Lambda/f \approx 1$~TeV for
$30<M_{e^*}<100$ GeV. 
Similar sensitivity is evident from the
excited quark search, for $q^{*}$ with dominantly electroweak couplings 
($f_s=0$).

It is possible to set mass limits on excited fermions from the
limit curves on $f/\Lambda $ versus $M_{f^{*}}$ if one makes the further
assumption that $f/\Lambda =1/M_{f^{*}}$.\footnote{
In the Lagrangian convention of the Particle Data Group \cite{PDG} for
excited fermion transitions, our mass limits correspond to $\lambda
_\gamma >1$ for excited electrons, $\lambda _W>1/\sqrt{2}$ for excited
neutrinos, and $\lambda _\gamma >\frac 23$ for excited $u$ quarks.}
For this case, excited electrons
are ruled out at the 95\% confidence level in the mass interval 30--200~GeV
using the combined limit from all three decay modes. Excited neutrinos are
excluded over the range 40--96~GeV.  Excited quarks with only
electroweak coupling are excluded over the range 40--169~GeV using the
combined limit from 
$q^{*}\rightarrow q\,\gamma $ and $q^{*}\rightarrow q\,W$.

The LEP experiments have recently reported searches \cite{LEPonefive,LEPtwo}
for excited leptons operating with center-of-mass energies $\sqrt{s}=161$
GeV and $\sqrt{s}=130$--$140$~GeV (LEP upper limits on $\lambda
/M_{l^{*}}$ must be multiplied by $\sqrt{2}$ to be
compared with our limits on $f/\Lambda$). The LEP single and pair
production upper limits on 
excited electrons ($f=f^{\prime }$) and excited electron neutrinos ($
f=-f^{\prime }$) are shown by the shaded lines in Figs.~\ref{fig:estar_flam}
and \ref{fig:nstar_flam}. The region above and to the left of the lines is
excluded at the 95\% confidence level. The sensitivity to excited leptons
above the pair production threshold and below 160--170 GeV is slightly better
than reported here, but the present analysis extends these limits to
well beyond 
170~GeV for excited electrons.

\section{Summary}

\label{sec:conclusion}

We have searched for heavy excited states of electrons, neutrinos, and
quarks using
9.4~pb$^{-1}$ of $e^{+}p$ collisions at a center-of-mass energy of 300~GeV
recorded with the ZEUS detector at HERA. No evidence of a signal was found
in any of eight distinct decay topologies. Upper limits at the 95\% confidence
level are derived on the production cross section times branching ratio
which are typically an order of magnitude better than our previous limits
from $e^{-}p$ collisions. We also set upper limits on the 
coupling  $f/\Lambda $ as a function of the excited fermion mass.  
With the choice $f/\Lambda
=1/M_{f^{*}}$, we exclude excited electrons with mass between 30 and  200~GeV,
excited electron neutrinos with mass between 40 and  96~GeV, and excited quarks
coupled electroweakly with mass between 40 and  169~GeV.

\section*{Acknowledgements}
We appreciate the contributions to the construction and maintenance of
the ZEUS detector by many people who are not listed as authors.  We
especially thank the DESY computing staff for providing the data analysis
environment and the HERA machine group for their
outstanding operation of the collider.  Finally, we thank the DESY
directorate for strong support and encouragement.

\section*{Appendix}

\label{sec:limitproc}

The procedure used for calculating the upper limit curves on excited fermion
(of mass, $M_f^{*}$) production is obtained from the spectrum in 
observed mass, $M$.  The formulation can be derived starting with the 
Poisson likelihood for observing $n$ events: 
\begin{equation}
{\cal L}(\mu_s;\,n,\mu_b) \, = \, {\frac{1}{n!}}\, e^{-(\mu_s+\mu_b)} \,
(\mu_s+\mu_b)^n  \label{eq:poisson}
\end{equation}
Here $\mu_s$ is the average number of expected signal events and 
$\mu_b$ is the average number of expected background events. 
The upper limit on $\mu_s$ at the 95\%
confidence level, $N_{{\rm U.L.}}$, is obtained by solving the
integral equation: 
\begin{equation}
\int_0^{N_{{\rm U.L.}}} {\cal L}(\mu_s;\,n,\mu_b)\, d\mu_s \, = \, 0.95\,
\int_0^{\infty} {\cal L}(\mu_s;\,n,\mu_b)\, d\mu_s  \label{eq:confidence}
\end{equation}
The solution to Eq.~(\ref{eq:confidence}) using the likelihood of 
Eq.~(\ref{eq:poisson}) reduces to the convention recommended by the 
Particle Data Group \cite{PDGtwo} for setting upper limits in the 
presence of background.

For the application relevant to this work, we divide the
total number of observed events, $n$, into a spectrum in  
observed mass, $M$.  The bin size, $dM$, is chosen small enough so that there 
is at most one event in any single bin.   
The likelihood function corresponding to a
signal with mass $M_{f^*}$ is then the product of the likelihoods from each
bin.  This likelihood reduces to 
\begin{equation}
{\cal L}(M_{f^*}) \, = \, \exp[-(\mu_{s}(M_{f^*})+\mu_{b})]\; \prod_{j\in 
{\rm events}}^{n} (\mu_{s,j}(M_{f^*})+\mu_{b,j})  \label{eq:liketwo}
\end{equation}
where $\mu_s(M_{f^*})$ and $\mu_b$ are the total number of 
expected signal and background events, respectively. The signal
and background populations in each bin are given by 
\begin{eqnarray}
\mu_{s,j}(M_{f^*}) & = & G(M_{f^*},M)\bigg|_{M=M_j} dM \\
\mu_{b,j} & = & B(M)\bigg|_{M=M_j} dM
\end{eqnarray}
where $M_j$ is the mass corresponding to observed event $j$.
The differential background contribution $B(M)$ is obtained from a fit
to the Monte Carlo calculated spectrum. The signal
contribution expected from the experiment is assumed to 
have a Gaussian lineshape, $G(M_{f^*},M)$, 
with peak values and
widths parameterized as a function of $M_{f^*}$.  These are obtained 
from Gaussian fits to Monte Carlo generated mass spectra.\footnote{%
These spectra include experimental systematic and resolution 
effects only.  The intrinsic lineshape of the excited fermion is assumed 
small (narrow width approximation).}

Inserting these parameterizations into Eq.~(\ref{eq:liketwo}) provides 
a likelihood (up to multiplication by bin size) 
at each value of excited fermion mass, $M_{f^*}$.  Substituting into 
Eq.~(\ref{eq:confidence}), we observe that
the bin size, $dM$, cancels on both sides of the equation. 
The solution to this 
integral equation provides at each $M_{f^*}$ the value for 
$N_{{\rm U.L.}}$, the 95\% confidence limit on the number of events.

It is straightforward to generalize this upper limit determination for the
case when two decay modes of the same excited fermion are
considered. 
The overall likelihood is the product of the individual 
likelihoods for each separate decay chain: 
\begin{eqnarray}
\lefteqn{{\cal L}(M_{f^*})\, =\,{\cal L}_1(M_{f^*})\,{\cal L}_2(M_{f^*})\,
=\, \exp[-(\mu_s(M_{f^*})+\mu_b^{(1)}+\mu_b^{(2)})] }  \label{eq:twodecay} \\
& & \times \prod_{i\in {\rm events}}^{n_1} \Bigg[\Big({\frac{A_1}{A_1+A_2}}%
\Big)\mu_{s,i}(M_{f^*})+\mu_{b,i}^{(1)}\Bigg] \prod_{j\in {\rm events}%
}^{n_2} \Bigg[\Big({\frac{A_2}{A_1+A_2}}\Big)\mu_{s,j}(M_{f^*})+%
\mu_{b,j}^{(2)}\Bigg]  \nonumber
\end{eqnarray}
where $\mu_s(M_{f^*}) = \mu_s^{(1)}+\mu_s^{(2)}$ and $A_i$ is the acceptance
of the $i$th decay mode.  The procedure for more than
two decay chains is a trivial extension of this equation.

A systematic error from uncertainties in the acceptance and luminosity
is included into the upper limit calculation by convoluting a
Gaussian with the Poisson likelihood: 
\begin{equation}
{\cal L}(\mu_s(M_{f^*});\,n,\mu_b) \rightarrow \int_0^{\infty}
 {\frac{d\gamma}{\sqrt{%
2\pi}\delta}}\, \exp[-(\gamma-1)^2/2\delta^2]\; {\cal L}(\gamma%
\mu_s(M_{f^*});\,n,\gamma\mu_b)  \label{eq:convolute}
\end{equation}
where $\delta$, the fractional systematic uncertainty, is taken to be 14\%
for the searches reported here.
For the specific case when zero events are observed, this convolution
increases $N_{\rm U.L.}$ from 3.0 to 3.2.

\clearpage

\begin{table}[p]
\centerline{
\begin{tabular}{|l|l|ccc|}
\hline
Channel & Signature & $A$  & $N_{\rm obs}$ & $N_{\rm exp}$  \\ 
\hline
\begin{tabular}{l} $e^*\rightarrow e\,\gamma$ \\ \end{tabular} &
\begin{tabular}{l} 2 EM clusters \\ \end{tabular} &
75\% & 103 & 
\begin{tabular}{c} $97.5\pm 4.8$ \\ ({\tiny COMPTON}: 81.5) \\
({\tiny NC DIS}: 16.0) \\ \end{tabular} 
\\ \hline
\begin{tabular}{l}
$e^* \rightarrow e\, Z^0 \rightarrow e\,q\,\bar q$ \\ 
$\nu^*_{\rm e} \rightarrow e\, W \rightarrow e\,q\,\bar q$ \\
\end{tabular} &
\begin{tabular}{l} EM cluster \\ 
large $E_{\rm t,had}$ and $M_{\rm had}$ \\ \end{tabular} &
\begin{tabular}{c} 55\% \\ 70\% \\ \end{tabular} & 
21 & 
\begin{tabular}{c} $15.3\pm1.5$ \\ ({\tiny NC DIS}) \\ \end{tabular}
\\ \hline
\begin{tabular}{l}
$e^* \rightarrow \nu_{\rm e}\, W \rightarrow \nu_{\rm e}\, q\, \bar q $ \\
$\nu^*_{\rm e} \rightarrow \nu_{\rm e}\, Z^0 \rightarrow \nu_{\rm e}\, q\,\bar q $ \\
\end{tabular} & 
\begin{tabular}{l} \PT \\ 
large $E_{\rm t,had}$ and $M_{\rm had}$ \\ \end{tabular} &
\begin{tabular}{c} 65\% \\ 75\% \\ \end{tabular}
& 13 
& \begin{tabular}{c} $8.4\pm1.5$ \\ ({\tiny CC DIS}: 4.8) \\
({\tiny NC DIS}: 1.0) \\ ({\tiny PHP}: 2.6) \\ \end{tabular}
\\ \hline
\begin{tabular}{l}
$\nu^*_{\rm e}\rightarrow \nu_{\rm e}\,\gamma$
\end{tabular} & 
\begin{tabular}{l} \PT \\ EM cluster (no track) \\ \end{tabular} &
50\% & 0 
& \begin{tabular}{c} $0.4\pm0.1$ \\ ({\tiny CC DIS}) \end{tabular}
\\  \hline
%
%
\begin{tabular}{l}
$e^* \rightarrow e\,Z^0 \rightarrow e\,\nu\,\bar\nu$ \\
$e^* \rightarrow \nu_{\rm e}\, W \rightarrow e\,\nu_{\rm
e}\,\bar\nu_{\rm e}$ \\
\end{tabular} &
\begin{tabular}{l} \PT \\ EM cluster (track) \\ \end{tabular}
& \begin{tabular}{c}70\% \\ 60\% \\ \end{tabular} 
& 1 
& \begin{tabular}{c} 1.0$\pm$0.2 \\ ({\tiny CC DIS}: 0.1) \\
({\tiny NC DIS}: 0.45) \\ ({\tiny W PROD'N}: 0.45) \\ \end{tabular}
\\ \hline
\begin{tabular}{l}
$\nu^*_{\rm e} \rightarrow e\, W \rightarrow e\, \overline{e}\,\nu_{\rm e}$ \\
\end{tabular} &
\begin{tabular}{l} \PT \\ 2 EM clusters \\ \end{tabular} 
& 60\% & 0 
& \begin{tabular}{c} $\approx 0$ \\ \end{tabular}
\\ \hline
\begin{tabular}{l}
$q^*\rightarrow q\,\gamma$
\end{tabular} &
\begin{tabular}{l} EM cluster (no track) \\ large $E_{\rm t,had}$ \\ \end{tabular} 
& 60\% & 18 
& \begin{tabular}{c} $23.5\pm2.5$ \\ ({{\tiny PROMPT} $\gamma$}: 14.4) 
\\ ({\tiny PHP}: 5.1) \\ ({\tiny NC DIS}: 4.0) \end{tabular}
\\ \hline
\begin{tabular}{l}
$q^* \rightarrow q\, W \rightarrow q\, \overline{e}\,\nu_{\rm e} $ \\ 
\end{tabular} &
\begin{tabular}{l} \PT \\ EM cluster \\ large $E_{\rm t,had}$  \\ \end{tabular}
& 50\% & 0 & \begin{tabular}{c} 1.5$\pm$0.3 \\ ({\tiny CC DIS}: 0.5) \\
({\tiny NC DIS}: 0.75) \\ ({\tiny W PROD'N}: 0.2) \end{tabular} 
\\ \hline
\end{tabular}
}
\caption{\label{tab:effic}
The $f^{*}$ decay signature, 
typical acceptance ($A$), 
number of observed events ($N_{\rm obs}$), 
and number of expected background events ($N_{\rm exp}$)
for the excited fermion decay
topologies investigated.
}
\end{table}

\clearpage


\end{document}